\documentclass[onecolumn]{aa}

\usepackage[T1]{fontenc}
\usepackage[latin1]{inputenc}
\usepackage{wasysym}
\usepackage{graphicx}
\usepackage{natbib}
\usepackage{txfonts}

\newcommand{\eq}[1]{\begin{equation}  #1 \end{equation}}

\newcommand{\eqa}[1]{\begin{eqnarray}   #1 \end{eqnarray}}

\newcommand{\br}[1]{\left( #1 \right)}
\newcommand{\bc}[1]{\left\{ #1 \right\}}
\newcommand{\bb}[1]{\left[ #1 \right]}
\newcommand{\ba}[1]{\left\langle #1 \right\rangle}

\newcommand{\nn}{\nonumber}

\newcommand{\dd}{{\rm d}}
\newcommand{\expo}[1]{~{\rm e}^{ #1 }}
\newcommand{\vek}[1]{\mbox{\boldmath $#1$}}
\newcommand{\svek}[1]{\mbox{\boldmath \scriptsize $#1$}}  
\newcommand{\ic}{{\rm i}}

\newcommand{\wigner}[6]{\left( \begin{array}{ccc} #1 & #2 & #3\\#4 & #5 & #6 \end{array} \right)}

\begin{document}

\title{Bispectrum covariance in the flat-sky limit}

\author{B. Joachimi\inst{1,2} \and X. Shi\inst{1} \and P. Schneider\inst{1}}

\offprints{B. Joachimi,\\
    \email{joachimi@astro.uni-bonn.de}
}

\institute{Argelander-Institut f\"ur Astronomie (AIfA), Universit\"at Bonn, Auf dem H\"ugel 71, 53121 Bonn, Germany
\and
Department of Physics and Astronomy, University College London, London WC1E 6BT, UK
}

\date{Received 16 July 2009 / Accepted 20 October 2009}

\abstract{}
{To probe cosmological fields beyond the Gaussian level, three-point statistics can be used, all of which are related to the bispectrum. Hence, measurements of CMB anisotropies, galaxy clustering, and weak gravitational lensing alike have to rely upon an accurate theoretical background concerning the bispectrum and its noise properties. If only small portions of the sky are considered, it is often desirable to perform the analysis in the flat-sky limit. We aim at a formal, detailed derivation of the bispectrum covariance in the flat-sky approximation, focusing on a pure two-dimensional Fourier-plane approach.}
{We define an unbiased estimator of the bispectrum, which takes the average over the overlap of annuli in Fourier space, and compute its full covariance. The outcome of our formalism is compared to the flat-sky spherical harmonic approximation in terms of the covariance, the behavior under parity transformations, and the information content. We introduce a geometrical interpretation of the averaging process in the estimator, thus providing an intuitive understanding.}
{Contrary to foregoing work, we find a difference by a factor of two between the covariances of the Fourier-plane and the spherical harmonic approach. We argue that this discrepancy can be explained by the differing behavior with respect to parity. However, in an exemplary analysis it is demonstrated that the Fisher information of both formalisms agrees to high accuracy. Via the geometrical interpretation we are able to link the normalization in the bispectrum estimator to the area enclosed by the triangle configuration at consideration as well as to the Wigner symbol, which leads to convenient approximation formulae for the covariances of both approaches.}
{}


\maketitle

\section{Introduction}
\label{sec:introduction}

As the concordance model of cosmology becomes more and more consolidated, the focus increasingly turns towards probing effects beyond the standard paradigm, such as non-Gaussian initial conditions or the evolution of the large-scale structure in the highly non-linear regime. To lowest order, these effects can be measured by three-point statistics of the underlying fields, all of which are related to the bispectrum. Hence, work in both theory and observations concerning the bispectrum and its noise properties has been undertaken for CMB measurements \citep[e.g.][]{hu00,cooray08}, galaxy clustering surveys \citep[e.g.][]{scoccimarro00,scoccimarro01a,sefusatti06}, or, more recently, weak gravitational lensing on cosmological scales \citep[e.g.][]{bernardeau02a,jarvis04,takada04}.

While theoretical computations at the bispectrum level are already considerably more demanding than for second-order statistics, this does apply even more so to the bispectrum covariance, which is a six-point statistic. On the full sky calculations are done by expanding the signal into spherical harmonics. If only small angular scales are considered, it is often more convenient to use a flat-sky approximation and work in terms of Fourier amplitudes. In the case of weak lensing the flat-sky limit is appropriate for practically all applications because signal correlations can only be measured up to separations of a few degrees.

Although other approaches exist in the literature \citep[e.g.][]{matarrese97,sefusatti06}, a lot of work is done within a flat-sky spherical harmonic formalism \citep{hu00}, which suffers -- at least formally -- from drawbacks. For instance, the resulting flat-sky expressions are valid only for integer arguments and thus for a bin width of unity, whereas it is desirable to evaluate the bispectrum and its covariance at real-valued angular frequencies and e.g. a logarithmic binning. The formulae still contain Wigner symbols whose physical meaning within a flat-sky consideration remain obscure. As the spherical harmonic expansion can only be done on the full unit sphere, the finite size of the survey at consideration is usually accounted for by multiplying a factor, containing the sky coverage, by hand. Moreover, the accuracy of some of the approximations in the transition between full sky and two-dimensional plane \citep[see][]{hu00} is uncertain.

This work aims at clarifying the derivation of bispectrum covariances in the flat-sky limit. We attempt to do so by presenting a detailed calculation which is purely based on the two-dimensional Fourier formalism, followed by a comparison of this approach with the flat-sky spherical harmonic results in terms of their covariance, the behavior under parity transformations, and the information content. Moreover, we provide further insight and illustration by establishing relations between Wigner symbols, the averaging process in the bispectrum estimator, and a geometrical view.

The outline of this note is as follows: In Sect.$\,$\ref{sec:estimator} a bispectrum estimator is defined and shown to be unbiased. Section \ref{sec:averaging} introduces a geometrical interpretation, which is then applied to deal with the issue of degenerate triangle configurations. In Sect.$\,$\ref{sec:covariance} the covariance of the estimator defined beforehand is computed. The result is compared with the spherical harmonics approach and demonstrated to be equivalent in terms of information content in Sect.$\,$\ref{sec:equivalence}. To explain the differences between the covariances, we also discuss the treatment of parity in both formalisms. We summarize our findings and conclude in Sect.$\,$\ref{sec:conclusions}. To avoid confusion, we refrain from using the term \lq flat sky\rq\ in the following, but refer to our formalism as \lq Fourier-plane\rq\ and to the approach as e.g. given in \citet{hu00} as \lq spherical harmonic\rq (both are flat-sky approximations).

\section{Bispectrum estimator}
\label{sec:estimator}

We consider a continuous, two-dimensional random field $g$ with mean zero, which is characterized by its complex Fourier amplitudes $g(\vek{\ell})$, where $\vek{\ell}$ denotes the angular frequency vector. Throughout, it will be assumed that this field is statistically homogeneous, i.e. invariant under translations, and statistically isotropic, i.e. invariant under rotations. In a cosmological context $g$ could for instance represent the temperature fluctuations of the CMB, the number density contrast of galaxy surveys, or the weak lensing convergence.

In what follows we will largely follow the approach of \citet{joachimi08}, assuming likewise measurements in a compact, contiguous survey of size $A$. We will restrict our considerations to an angular extent much smaller than the size of the survey, i.e. to $\ell \gg \pi/\theta_{\rm max}$, where $\theta_{\rm max}$ is the maximum separation allowed by the survey geometry. Boundary effects due to the finite field size, as e.g. discussed in \citet{joachimi08} for the second-order level, can then be safely neglected.

Furthermore, we will not explicitly consider additional noise terms due to the discrete sampling of the continuous field $g$, for ease of notation. To account for these shot noise or, in the case of weak lensing, shape noise terms in the covariance, they can simply be added to the second-order measures, so in this Fourier space approach, to the power spectra \citep[e.g.][]{kaiser98,hu99}. 

Note that the galaxy ellipticity, and not the convergence $\kappa$, is the direct observable in weak lensing. However, in absence of shape noise and for $\ell \gg 1$, the estimators in terms of the galaxy ellipticity, as given in \citet{joachimi08}, can be re-written directly in terms of $\kappa$. Thus, without loss of generality, one can consider the convergence as the observable that the estimator is based on.

For a statistically homogeneous and isotropic random field one defines the bispectrum as
\eq{
\label{eq:defbispectrum}
\ba{g(\vek{\ell}_1)\; g(\vek{\ell}_2)\; g(\vek{\ell}_3)} = (2\pi)^2\; \delta^{(2)}_{\rm D}(\vek{\ell}_1+\vek{\ell}_2+\vek{\ell}_3)\; B(\ell_1,\ell_2,\ell_3)\;,
}
where $\delta^{(2)}_{\rm D}(\vek{\ell})$ is the two-dimensional Dirac delta-distribution. It ensures in (\ref{eq:defbispectrum}) that the three angular frequency vectors form a triangle. For the assumed properties of $g$ the bispectrum has three independent components, for which we have chosen the triangle side lengths $|\vek{\ell}_i| \equiv \ell_i$. For the absolute values of $\vek{\ell}$ the triangle condition translates into the requirement $|\ell_1 - \ell_2| \leq \ell_3 \leq \ell_1 + \ell_2$ or equivalently for any permutation of the $\ell_i$.

Similarly to \citet{joachimi08}, we construct an estimator of the bispectrum by averaging configurations over annuli, where here one has the complication of allowing only those combinations of angular frequency vectors that form a triangle. The area of an annulus with mean radius $\bar{\ell}_i$ is given by
\eq{
\label{eq:annuli}
A_R(\bar{\ell}_i) = 2\pi \bar{\ell}_i \Delta \ell_i
}
with the bin size $\Delta \ell_i$. Then we define the estimator
\eq{
\label{eq:estimator}
\hat{B}(\bar{\ell}_1,\bar{\ell}_2,\bar{\ell}_3) := \frac{(2\pi)^2}{A} \Lambda^{-1} \br{\bar{\ell}_1,\bar{\ell}_2,\bar{\ell}_3} \int_{A_R(\bar{\ell}_1)} \frac{ \dd^2 \ell_1}{A_R(\bar{\ell}_1)} \int_{A_R(\bar{\ell}_2)} \frac{ \dd^2 \ell_2}{A_R(\bar{\ell}_2)} \int_{A_R(\bar{\ell}_3)} \frac{ \dd^2 \ell_3}{A_R(\bar{\ell}_3)}\; \delta^{(2)}_{\rm D}(\vek{\ell}_1+\vek{\ell}_2+\vek{\ell}_3)\; g(\vek{\ell}_1)\; g(\vek{\ell}_2)\; g(\vek{\ell}_3)\;,
}
where $\Lambda$ is a function that is related to the fraction of angular frequency combinations allowed by the triangle condition. It is defined such that (\ref{eq:estimator}) is unbiased, its explicit form being calculated below. Note that this bispectrum estimator is invariant under any permutation of its arguments since $\Lambda$ is symmetric as will be shown below.

In the following, we demonstrate that (\ref{eq:estimator}) is unbiased by computing the ensemble average,
\eqa{
\label{eq:ensembleaverage}
\ba{\hat{B}(\bar{\ell}_1,\bar{\ell}_2,\bar{\ell}_3)}  &=&  \frac{(2\pi)^2}{A} \Lambda^{-1} \br{\bar{\ell}_1,\bar{\ell}_2,\bar{\ell}_3} \int_{A_R(\bar{\ell}_1)} \frac{ \dd^2 \ell_1}{A_R(\bar{\ell}_1)} \int_{A_R(\bar{\ell}_2)} \frac{ \dd^2 \ell_2}{A_R(\bar{\ell}_2)} \int_{A_R(\bar{\ell}_3)} \frac{ \dd^2 \ell_3}{A_R(\bar{\ell}_3)}\; (2\pi)^2 \br{ \delta^{(2)}_{\rm D}(\vek{\ell}_1+\vek{\ell}_2+\vek{\ell}_3) }^2 B(\ell_1,\ell_2,\ell_3)\\ \nn
&=& (2\pi)^2 \Lambda^{-1} \br{\bar{\ell}_1,\bar{\ell}_2,\bar{\ell}_3} \int_{A_R(\bar{\ell}_1)} \frac{ \dd^2 \ell_1}{A_R(\bar{\ell}_1)} \int_{A_R(\bar{\ell}_2)} \frac{ \dd^2 \ell_2}{A_R(\bar{\ell}_2)} \int_{A_R(\bar{\ell}_3)} \frac{ \dd^2 \ell_3}{A_R(\bar{\ell}_3)}\; \delta^{(2)}_{\rm D}(\vek{\ell}_1+\vek{\ell}_2+\vek{\ell}_3)\; B(\ell_1,\ell_2,\ell_3)\;.
}
In the first step the definition of the bispectrum (\ref{eq:estimator}) was inserted. The appearance of a squared delta-distribution requires taking into account the finite survey size. As shown in \citet{joachimi08}, one can identify
\eq{
\label{eq:delta0}
\br{ \delta^{(2)}_{\rm D}(\vek{\ell}) }^2 \rightarrow \frac{A}{(2\pi)^2}\; \delta^{(2)}_{\rm D}(\vek{\ell})\:,
}
which results in the second equality of (\ref{eq:ensembleaverage}). 

Since the bispectrum only depends on the magnitudes of the angular frequency vectors we can perform the integrations over the polar angles of the $\vek{\ell}$-integrals. If $\varphi_{\ell_i}$ denotes the polar angle of $\vek{\ell}_i$, one gets
\eqa{
\label{eq:gaunt2}
&& \int_0^{2\pi} \dd \varphi_{\ell_1} \int_0^{2\pi} \dd \varphi_{\ell_2} \int_0^{2\pi} \dd \varphi_{\ell_3}\; \delta^{(2)}_{\rm D}(\vek{\ell}_1+\vek{\ell}_2+\vek{\ell}_3) = \int_0^{2\pi} \dd \varphi_{\ell_1} \int_0^{2\pi} \dd \varphi_{\ell_2} \int_0^{2\pi} \dd \varphi_{\ell_3}\; \int \frac{\dd^2 \theta}{(2\pi)^2} \expo{\ic \br{\svek{\ell}_1 + \svek{\ell}_2 + \svek{\ell}_3} \cdot \svek{\theta}}\\ \nn
&=& \int \frac{\dd^2 \theta}{(2\pi)^2}\; (2\pi)^3 J_0(\ell_1 \theta)\; J_0(\ell_2 \theta)\; J_0(\ell_3 \theta) \; = \; (2\pi)^2 \int \dd \theta\; \theta\; J_0(\ell_1 \theta)\; J_0(\ell_2 \theta)\; J_0(\ell_3 \theta) \; = \; 2\pi\; \Lambda \br{\ell_1,\ell_2,\ell_3}\;.
}
After inserting one possible representation of the delta-distribution in the first equality, we have made use of the definition of the Bessel function of the first kind of order 0,
\eq{
\label{eq:defbessel}
J_0(x) = \int_0^{2\pi} \frac{\dd \varphi}{2\pi} \expo{\ic x \cos \varphi}\;.
}
The result of the integral over three Bessel functions is taken from \citet{gradshteyn}, formula no. 6.578.9, where we have defined
\eq{
\label{eq:deflambda}
\Lambda \br{\ell_1,\ell_2,\ell_3} \equiv \left\{  \begin{array}{ll} \bc {\frac{1}{4} \sqrt{2 \ell_1^2 \ell_2^2 + 2 \ell_1^2 \ell_3^2 + 2 \ell_2^2 \ell_3^2 - \ell_1^4 - \ell_2^4 - \ell_3^4} }^{-1} & \mbox{if}~~~ |\ell_1-\ell_2| < \ell_3 < \ell_1 + \ell_2\\ 0 & \mbox{else}\\ \end{array} \right.\;,
}
i.e. if $\ell_1,\ell_2,\ell_3$ are chosen such that they can form the sides of a triangle, then $\Lambda^{-1}$ is the area of this triangle. Hence, (\ref{eq:gaunt2}) represents the defining equation for $\Lambda$. The set of integrations (\ref{eq:gaunt2}) is also performed within the spherical harmonic approach, see the appendix of \citet{hu00}, with a different result, which will be investigated in Sect.$\,$\ref{sec:comp}. Note furthermore that $\Lambda \br{\ell_1,\ell_2,\ell_3}=0$ in case the angular frequency vectors are collinear or equivalently, if $\ell_i + \ell_j = \ell_k$ for some combination $i,j,k \in \bc{1,2,3}$. At the same time, the bispectrum is non-zero for these degenerate triangle configurations, see (\ref{eq:defbispectrum}). For the time being, we exclude degenerate triangles from the derivation, but develop a treatment for these cases in Sect.$\,$\ref{sec:collapsed}.

Inserting (\ref{eq:gaunt2}) into (\ref{eq:ensembleaverage}), one obtains
\eq{
\label{eq:ensembleaverage2}
\ba{\hat{B}(\bar{\ell}_1,\bar{\ell}_2,\bar{\ell}_3)} = (2\pi)^3 \Lambda^{-1}  \br{\bar{\ell}_1,\bar{\ell}_2,\bar{\ell}_3} \int_{\bar{\ell}_1-1/2 \Delta \ell}^{\bar{\ell}_1+1/2 \Delta \ell} \frac{ \dd \ell_1 \ell_1}{A_R(\bar{\ell}_1)} \int_{\bar{\ell}_2-1/2 \Delta \ell}^{\bar{\ell}_2+1/2 \Delta \ell} \frac{ \dd \ell_2 \ell_2}{A_R(\bar{\ell}_2)} \int_{\bar{\ell}_3-1/2 \Delta \ell}^{\bar{\ell}_3+1/2 \Delta \ell} \frac{ \dd \ell_3 \ell_3}{A_R(\bar{\ell}_3)}\; \Lambda \br{\ell_1,\ell_2,\ell_3} B(\ell_1,\ell_2,\ell_3)\;.
}
Analogous to the derivation at the level of second-order statistics \citep{joachimi08} we assume now that the annuli are thin enough such that $\Lambda$ within the integral, evaluated at the average $\ell$-values, can be taken out of the integration. Applying in addition (\ref{eq:annuli}), one arrives at
\eq{
\label{eq:ensembleaverage3}
\ba{\hat{B}(\bar{\ell}_1,\bar{\ell}_2,\bar{\ell}_3)} \approx  \int_{\bar{\ell}_1-1/2 \Delta \ell}^{\bar{\ell}_1+1/2 \Delta \ell} \frac{ \dd \ell_1 \ell_1}{\bar{\ell}_1 \Delta \ell_1} \int_{\bar{\ell}_2-1/2 \Delta \ell}^{\bar{\ell}_2+1/2 \Delta \ell} \frac{ \dd \ell_2 \ell_2}{\bar{\ell}_2 \Delta \ell_2} \int_{\bar{\ell}_3-1/2 \Delta \ell}^{\bar{\ell}_3+1/2 \Delta \ell} \frac{ \dd \ell_3 \ell_3}{\bar{\ell}_3 \Delta \ell_3}\; B(\ell_1,\ell_2,\ell_3) \equiv B(\bar{\ell}_1,\bar{\ell}_2,\bar{\ell}_3)\;,
}
where in the last step the bin-averaged bispectrum was defined. Hence, (\ref{eq:estimator}) defines an unbiased estimator of the bispectrum. Following the restrictions on (\ref{eq:gaunt2}), this estimator is non-zero if the condition $|\bar{\ell}_1 - \bar{\ell}_2| < \bar{\ell}_3 < \bar{\ell}_1 + \bar{\ell}_2$, or likewise for all permutations, holds.

\section{Averaging over triangles}
\label{sec:averaging}

A central step in the construction of the bispectrum estimator (\ref{eq:estimator}) is the correct treatment of the averaging over annuli, given the triangle condition. This section provides an illustrative, geometrical interpretation of the averaging process and applies this view to a practical treatment of degenerate triangle configurations.

\subsection{Geometrical interpretation}

\begin{figure}[t]
\begin{minipage}[c]{.7\textwidth}
\centering
\includegraphics[scale=.23]{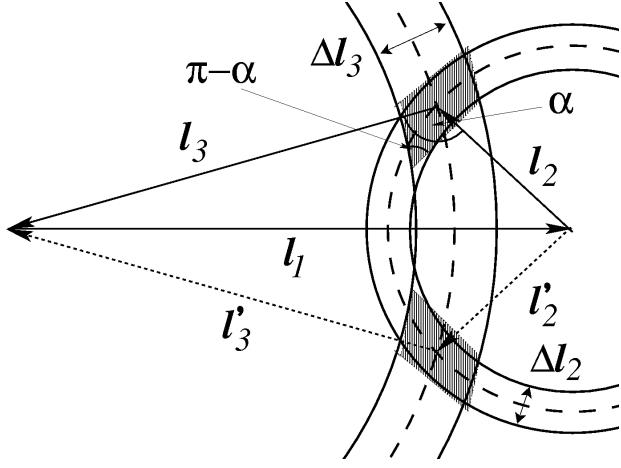}
\end{minipage}%
\begin{minipage}[c]{.3\textwidth}
\caption{Sketch of the annuli and their overlap for fixed $\vek{\ell}_1$. The region of overlap is approximated by the shaded parallelograms. Note that due to mirror symmetry a second shaded area, related to the triangle $\vek{\ell}_1$, $\vek{\ell}'_2$, $\vek{\ell}'_3$, contributes as well.}
\label{fig:sketch_overlap}
\end{minipage}
\end{figure}

Without loss of generality consider $\vek{\ell}_1$ to be fixed. Due to the assumed statistical isotropy of the underlying random field the angular integration over $\varphi_{\ell_1}$ is expected to simply reduce to an average over all directions of $\vek{\ell}_1$. Then the geometric situation in the Fourier plane can be seen as in Fig.$\,$\ref{fig:sketch_overlap}. For a given triangle, composed of the mean vectors $\vek{\ell}_1$, $\vek{\ell}_2$, $\vek{\ell}_3$ with lengths $\bar{\ell}_i$ for $i=\bc{1,2,3}$, the annuli for $\bar{\ell}_2$ and $\bar{\ell}_3$ are shown. Due to the triangle condition, the average is not taken over the whole area of the annuli, but merely over the region that the annuli have in common. This area of overlap is well approximated by a parallelogram of size $A_\parallel=\Delta \ell_2 \Delta \ell_3 / \sin \alpha$, where $\alpha$ is the internal angle of the triangle opposite $\bar{\ell}_1$. This relation can readily be computed from the geometry of the sketch and by noting $\sin \alpha = \sin (\pi-\alpha)$.

The configuration is mirror-symmetric with respect to an axis through $\vek{\ell}_1$. Correspondingly, another area of overlap of the same size, which is connected to the triangle $\vek{\ell}_1$, $\vek{\ell}'_2$, $\vek{\ell}'_3$, contributes as well. Noting that axis reflection is in two dimensions equivalent to the parity transformation, the averaging is performed over triangles of both parities. A detailed discussion on this issue is given in Sect.$\,$\ref{sec:parity}.

As the angle $\alpha$ can also be related to the size of the triangle at consideration, $\Lambda^{-1} = (1/2)\, \bar{\ell}_2 \bar{\ell}_3 \sin \alpha$, one finds the following correspondence of expressions: 
\eq{
\label{eq:geometriccorrespondence}
\int_0^{2\pi} \frac{\dd \varphi_{\ell_1}}{2\pi} \int_0^{2\pi} \frac{\dd \varphi_{\ell_2}}{2\pi} \int_0^{2\pi} \frac{\dd \varphi_{\ell_3}}{2\pi}\; \delta^{(2)}_{\rm D}(\vek{\ell}_1+\vek{\ell}_2+\vek{\ell}_3) \; = \; \br{ 2\pi }^{-2}\; \Lambda \br{\bar{\ell}_1,\bar{\ell}_2,\bar{\ell}_3} \; = \; \frac{2\, A_\parallel}{A_R(\bar{\ell}_2)\;A_R(\bar{\ell}_3)}\;,
}
where the first equality is an immediate consequence of (\ref{eq:gaunt2}). To arrive at the last expression, we used (\ref{eq:annuli}). Hence, the angular integration over the delta-distribution yields the ratio of the area of overlap $A_{\rm overlap}$, approximated by $2\,A_\parallel$, and the product of the area of the annuli the $\ell$-integrations (excluding the fixed $\vek{\ell}_1$) run over. This ratio is in turn proportional to the inverse of the area of the triangle spanned by the angular frequency vectors. Therefore, by placing a prefactor of $\Lambda^{-1}$ in the estimator (\ref{eq:estimator}), one replaces the normalization by the area of the annuli with the effective area, over which the average is actually performed.

Two approximations are involved in this picture. First, the shaded regions in Fig.$\,$\ref{fig:sketch_overlap} are approximated as parallelograms, which is a good assumption if the angle, at which the two annuli intersect, does not become too small. Moreover, the narrower the annuli, the less discrepancy between the area of the parallelogram and the actual overlap is expected. If the triangle approaches the degenerate case, where $\bar{\ell}_2$ and $\bar{\ell}_3$ eventually come to lie on $\bar{\ell}_1$, the area of overlap attains a more complex shape. In particular, the correspondence to the area of the triangle, whose inverse is divergent, does not hold anymore. Second, reconsidering (\ref{eq:ensembleaverage2}), we have replaced the average of $\Lambda$ over triangle side lengths by $\Lambda$, evaluated at the average side lengths. This approximation similarly breaks down for thick annuli and configurations in which a small change in the length of an angular frequency vector causes a strong change in the size of the overlap region, as is the case near degeneracy.

\begin{figure}[t]
\centering
\includegraphics[scale=.65,angle=270]{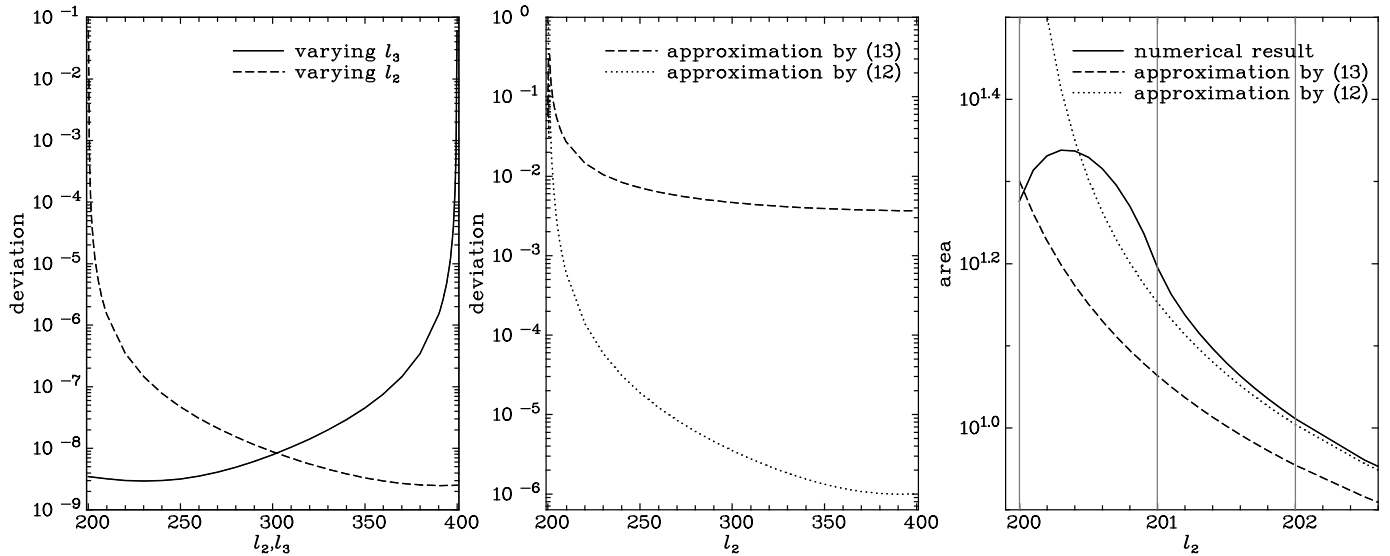}
\caption{Comparison of expressions for the overlap area of annuli. \textit{Left panel}: Relative deviation of (\ref{eq:areaproxy1}) from the overlap area of the annuli as a function of angular frequency. The bin width is kept constant at $\Delta=0.05$. The solid curve shows results for $\bar{\ell}_1,\bar{\ell}_2=200$ and varying $\bar{\ell}_3$, while the dashed curve corresponds to $\bar{\ell}_1=200,\bar{\ell}_3=400$ and varying $\bar{\ell}_2$. \textit{Center panel}: Same as above for the case $\bar{\ell}_1=200,\bar{\ell}_3=400$ and varying $\bar{\ell}_2$, but with $\Delta=1$. The dotted line illustrates the deviation of (\ref{eq:areaproxy1}), the dashed line the deviation of (\ref{eq:areaproxy2}). \textit{Right panel}: Area of overlap for the case $\bar{\ell}_1=200,\bar{\ell}_3=400$ and varying $\bar{\ell}_2$, with $\Delta=1$. The solid curve corresponds to the actual area, the dotted curve to (\ref{eq:areaproxy1}), and the dashed curve to (\ref{eq:areaproxy2}). Note that due to $\Delta=1$ only the values at integer values of $\bar{\ell}_2$ are relevant for the covariance calculation.}
\label{fig:triangleareas}
\end{figure}

In Fig.$\,$\ref{fig:triangleareas}, we have plotted the relative deviation of
\eq{
\label{eq:areaproxy1}
A_{\rm overlap}^{\rm n.d.} = \bar{\ell}_2\; \bar{\ell}_3\; \Delta \ell_2\; \Delta \ell_3\; \Lambda\br{\bar{\ell}_1,\bar{\ell}_2,\bar{\ell}_3}
}
from the actual area of the overlap region, which we calculated numerically. For simplicity, we assume a constant bin width $\Delta \ell_2 = \Delta \ell_3 \equiv \Delta$ for all computations related to Fig.$\,$\ref{fig:triangleareas}. For a small bin width $\Delta = 0.05$, given integer steps in $\ell$, we find for the two configurations considered in the top panel that the approximation of the overlap area by parallelograms is excellent for the vast majority of triangle configurations. However, as expected, the deviation rises sharply when approaching the degenerate case. Changing to $\Delta = 1$, i.e. the maximum meaningful bin width in this setup, the relative deviation is larger, but still very small except for triangles close to degeneracy.

\subsection{Degenerate triangles}
\label{sec:collapsed}

As discussed in the foregoing section, the approximations made in the course of the construction of the bispectrum estimator break down for degenerate triangle configurations. Equation (\ref{eq:geometriccorrespondence}) becomes invalid, the inverse area of the triangle $\Lambda$ diverging. Yet, to be of practical use, it is necessary to extend the validity of (\ref{eq:estimator}) to the case of degenerate triangles. We do so by making use of the geometrical interpretation of the averaging process.

\begin{figure}[t]
\begin{minipage}[c]{.7\textwidth}
\centering
\includegraphics[scale=.25]{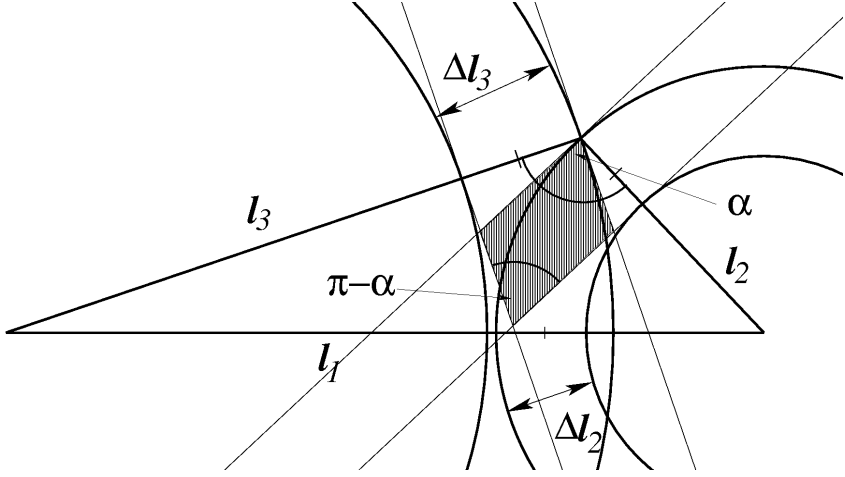}
\end{minipage}%
\begin{minipage}[c]{.3\textwidth}
\caption{Sketch of the region averaged over in case of a degenerate triangle, again for fixed $\vek{\ell}_1$. The depicted triangle has side lengths $\bar{\ell}_1$, $\bar{\ell}_2+\Delta \ell_2/2$, and $\bar{\ell}_3+\Delta \ell_3/2$. The shaded parallelogram approximates the region of overlap, the mirror-symmetric counterpart not being shown.}
\label{fig:sketch_collapsed}
\end{minipage}
\end{figure}

Still keeping $\vek{\ell}_1$ fixed, consider the situation of a degenerate triangle as sketched in Fig.$\,$\ref{fig:sketch_collapsed}. Here, $\bar{\ell}_1 = \bar{\ell}_2 + \bar{\ell}_3$, while the depicted triangle has side lengths $\bar{\ell}_1$, $\bar{\ell}_2+\Delta \ell_2/2$, and $\bar{\ell}_3+\Delta \ell_3/2$. Again, we identify a parallelogram that serves as an approximation for the overlap of the annuli, although, as the sketch suggests, with considerably lower accuracy. The relation between the internal angle $\alpha$ of the triangle to the internal angle of the parallelogram $\pi -\alpha$ holds as before, so that one can derive an analogous formula to (\ref{eq:areaproxy1}), but with modified triangle side lengths. Symmetrizing this argument for all three angular frequency vectors, we propose the following formula to compute the area of overlap in the degenerate case:
\eq{
\label{eq:areaproxy2}
A_{\rm overlap}^{\rm deg.} := \bar{\ell}_2\; \bar{\ell}_3\; \Delta \ell_2\; \Delta \ell_3\; \Lambda\br{\bar{\ell}_1 + \frac{\Delta \ell_1}{2},\bar{\ell}_2 + \frac{\Delta \ell_2}{2},\bar{\ell}_3 + \frac{\Delta \ell_3}{2}}\;.
}

As is evident from Fig.$\,$\ref{fig:triangleareas}, center panel, the relative deviation of (\ref{eq:areaproxy2}) from the true overlap area is still fairly small, but -- unsurprisingly -- noticeably stronger than for (\ref{eq:areaproxy1}). The right-hand panel gives the size of the overlap area for values of $\bar{\ell}_2$ close to 200, which is the degenerate case. Note that since this plot was determined for $\Delta =1$, the values relevant for the covariance calculation are only those at integer $\ell$. While the true overlap area curbs down to a finite value at $\bar{\ell}_2=200$, (\ref{eq:areaproxy1}) diverges. Still, for $\bar{\ell}_2=201$ it produces a fair and for $\bar{\ell}_2 \geq 202$ an excellent approximation. In the degenerate case (\ref{eq:areaproxy2}) is indeed capable of reproducing the size of the overlap area to good accuracy.

Thus, we suggest to incorporate degenerate triangle configurations into our formalism by replacing $\bar{\ell}_i \rightarrow \bar{\ell}_i + \Delta \ell_i/2$ in all arguments of $\Lambda$ for these cases. This way, we heuristically correct for the breakdown of approximations in the assignment of the actual area, over which triangle configurations are averaged. While the modification is at this stage only motivated by the geometrical interpretation, we will establish a more strict foundation of (\ref{eq:areaproxy2}) by relating it to Wigner symbols in Sect.$\,$\ref{sec:comp}.

\section{Bispectrum covariance}
\label{sec:covariance}

The covariance of the bispectrum is defined as
\eqa{
\label{eq:defcov}
{\rm Cov} \br{\hat{B}(\bar{\ell}_1,\bar{\ell}_2,\bar{\ell}_3),\; \hat{B}(\bar{\ell}_4,\bar{\ell}_5,\bar{\ell}_6)} &\equiv& \ba{ \br{ \hat{B}(\bar{\ell}_1,\bar{\ell}_2,\bar{\ell}_3) - \ba{\hat{B}(\bar{\ell}_1,\bar{\ell}_2,\bar{\ell}_3)}}\; \br{ \hat{B}(\bar{\ell}_4,\bar{\ell}_5,\bar{\ell}_6) - \ba{\hat{B}(\bar{\ell}_4,\bar{\ell}_5,\bar{\ell}_6)}} }\\ \nn
&=& \ba{ \hat{B}(\bar{\ell}_1,\bar{\ell}_2,\bar{\ell}_3)\; \hat{B}(\bar{\ell}_4,\bar{\ell}_5,\bar{\ell}_6) } - B(\bar{\ell}_1,\bar{\ell}_2,\bar{\ell}_3)\; B(\bar{\ell}_4,\bar{\ell}_5,\bar{\ell}_6)\;.
}
The computation of the correlator of two bispectrum estimators involves a 6-point correlator of $g$, which can be expanded into its connected parts as e.g. outlined in \citet{bernardeau02}. Denoting the connected correlators by a subscript c, which will only be done in this paragraph to avoid confusion, we obtain
\eqa{
\label{eq:expand6}
&& \ba{ g(\vek{\ell}_1)\; g(\vek{\ell}_2)\; g(\vek{\ell}_3)\; g(\vek{\ell}_4)\; g(\vek{\ell}_5)\; g(\vek{\ell}_6) }\\ \nn
&=& \ba{g(\vek{\ell}_1)\; g(\vek{\ell}_2)}_{\rm c} \ba{g(\vek{\ell}_3)\; g(\vek{\ell}_4)}_{\rm c} \ba{g(\vek{\ell}_5)\; g(\vek{\ell}_6)}_{\rm c} + \ba{g(\vek{\ell}_1)\; g(\vek{\ell}_2)}_{\rm c} \ba{g(\vek{\ell}_3)\; g(\vek{\ell}_5)}_{\rm c} \ba{g(\vek{\ell}_4)\; g(\vek{\ell}_6)}_{\rm c} + (\mbox{13 perm.})\\ \nn
&+& \ba{g(\vek{\ell}_1)\; g(\vek{\ell}_2)\; g(\vek{\ell}_3)}_{\rm c} \ba{g(\vek{\ell}_4)\; g(\vek{\ell}_5)\; g(\vek{\ell}_6)}_{\rm c} + \ba{g(\vek{\ell}_1)\; g(\vek{\ell}_2)\; g(\vek{\ell}_4)}_{\rm c} \ba{g(\vek{\ell}_3)\; g(\vek{\ell}_5)\; g(\vek{\ell}_6)}_{\rm c} + (\mbox{8 perm.})\\ \nn
&+& \ba{g(\vek{\ell}_1)\; g(\vek{\ell}_2)\; g(\vek{\ell}_3)\; g(\vek{\ell}_4)}_{\rm c} \ba{g(\vek{\ell}_5)\; g(\vek{\ell}_6)}_{\rm c} + \ba{g(\vek{\ell}_1)\; g(\vek{\ell}_2)\; g(\vek{\ell}_3)\; g(\vek{\ell}_5)}_{\rm c} \ba{g(\vek{\ell}_4)\; g(\vek{\ell}_6)}_{\rm c} + (\mbox{13 perm.})\\ \nn
&+& \ba{ g(\vek{\ell}_1)\; g(\vek{\ell}_2)\; g(\vek{\ell}_3)\; g(\vek{\ell}_4)\; g(\vek{\ell}_5)\; g(\vek{\ell}_6) }_{\rm c}\;,
}
where the permutations are to be taken with respect to the indices of the angular frequencies such that for each correlator, no combination of indices is repeated (as the individual correlators are invariant under permutations of the indices within that correlator). The resulting connected parts are related to spectra via
\eq{
\label{eq:defspectra}
\ba{ \prod_{i=1}^N g(\vek{\ell}_i) }_{\rm c} = (2\pi)^2\; \delta^{(2)}_{\rm D} \br{ \sum_{i=1}^N \vek{\ell}_i }\; P_N(\vek{\ell}_1,\,...\,,\vek{\ell}_N)\;,
}
where we identify $P_2(\vek{\ell}_1,\vek{\ell}_2) \equiv P(\ell_1)$ as the power spectrum and $P_3(\vek{\ell}_1,\vek{\ell}_2,\vek{\ell}_3) \equiv B(\ell_1,\ell_2,\ell_3)$ as the bispectrum. As the random field $g$ vanishes on average, $\ba{g(\vek{\ell})}=0$, only $P_4$ (the trispectrum) and $P_6$ (the pentaspectrum) will appear in addition in the covariance formula, see (\ref{eq:expand6}).

Introducing a shorthand notation $\int_{A_R(\bar{\ell}_i)} \dd^2 \ell_i / A_R(\bar{\ell}_i) \equiv \int_i$, one can write the correlator of the bispectrum estimators by using (\ref{eq:estimator}) as
\eqa{
\label{eq:cov_insertestimator}
\ba{ \hat{B}(\bar{\ell}_1,\bar{\ell}_2,\bar{\ell}_3)\; \hat{B}(\bar{\ell}_4,\bar{\ell}_5,\bar{\ell}_6) } &=& \frac{(2\,\pi)^4}{A^2}\; \Lambda^{-1} \br{\bar{\ell}_1,\bar{\ell}_2,\bar{\ell}_3} \Lambda^{-1} \br{\bar{\ell}_4,\bar{\ell}_5,\bar{\ell}_6} \int_1 \int_2 \int_3 \int_4 \int_5 \int_6\\ \nn
&& \hspace*{3cm} \times\; \delta^{(2)}_{\rm D}(\vek{\ell}_1+\vek{\ell}_2+\vek{\ell}_3)\; \delta^{(2)}_{\rm D}(\vek{\ell}_4+\vek{\ell}_5+\vek{\ell}_6)\; \ba{ g(\vek{\ell}_1)\; g(\vek{\ell}_2)\; g(\vek{\ell}_3)\; g(\vek{\ell}_4)\; g(\vek{\ell}_5)\; g(\vek{\ell}_6) }\;,
}
which then allows us to insert (\ref{eq:expand6}) and (\ref{eq:defspectra}). The resulting terms contain products of several delta-distribution. Concerning the terms containing three two-point correlators, one obtains e.g. 
\eqa{
\label{eq:deltacomb}
&& \delta^{(2)}_{\rm D}(\vek{\ell}_1+\vek{\ell}_2+\vek{\ell}_3)\; \delta^{(2)}_{\rm D}(\vek{\ell}_4+\vek{\ell}_5+\vek{\ell}_6)\; \delta^{(2)}_{\rm D}(\vek{\ell}_1+\vek{\ell}_2)\; \delta^{(2)}_{\rm D}(\vek{\ell}_3+\vek{\ell}_4)\; \delta^{(2)}_{\rm D}(\vek{\ell}_5+\vek{\ell}_6) P(\ell_1)\; P(\ell_3)\; P(\ell_5)\\ \nn
&=& \delta^{(2)}_{\rm D}(\vek{\ell}_3)\; \delta^{(2)}_{\rm D}(\vek{\ell}_4+\vek{\ell}_5+\vek{\ell}_6)\; \delta^{(2)}_{\rm D}(\vek{\ell}_1+\vek{\ell}_2)\; \delta^{(2)}_{\rm D}(\vek{\ell}_3+\vek{\ell}_4)\; \delta^{(2)}_{\rm D}(\vek{\ell}_5+\vek{\ell}_6) P(\ell_1)\; P(0)\; P(\ell_5) \; = \; 0\;,
}
and likewise for all other terms in which the correlators do not contain one angular frequency each out of the sets $\bc{\vek{\ell}_1,\vek{\ell}_2,\vek{\ell}_3}$ and $\bc{\vek{\ell}_4,\vek{\ell}_5,\vek{\ell}_6}$. A similar argument holds for the terms composed of power spectrum and trispectrum, where the trispectrum is readily shown to vanish if the two-point correlator contains both angular frequencies out of the same of the sets mentioned above. This way, the number of terms with three power spectra reduces to 6, the number of terms with trispectrum and power spectrum to 9.

To proceed, we demonstrate the treatment of some exemplary terms in the covariance, for instance
\eqa{
\label{eq:covcalc_2}
&& \int_1 \int_2 \int_3 \int_4 \int_5 \int_6 \delta^{(2)}_{\rm D}(\vek{\ell}_1+\vek{\ell}_2+\vek{\ell}_3)\; \delta^{(2)}_{\rm D}(\vek{\ell}_4+\vek{\ell}_5+\vek{\ell}_6)\; \ba{g(\vek{\ell}_1)\; g(\vek{\ell}_4)} \ba{g(\vek{\ell}_2)\; g(\vek{\ell}_5)} \ba{g(\vek{\ell}_3)\; g(\vek{\ell}_6)}\\ \nn
&=& (2\pi)^6 \int_1 \int_2 \int_3 \int_4 \int_5 \int_6 \delta^{(2)}_{\rm D}(\vek{\ell}_1+\vek{\ell}_2+\vek{\ell}_3)\; \delta^{(2)}_{\rm D}(\vek{\ell}_4+\vek{\ell}_5+\vek{\ell}_6)\; \delta^{(2)}_{\rm D}(\vek{\ell}_1+\vek{\ell}_4)\; \delta^{(2)}_{\rm D}(\vek{\ell}_2+\vek{\ell}_5)\; \delta^{(2)}_{\rm D}(\vek{\ell}_3+\vek{\ell}_6)\; P(\ell_1)\; P(\ell_2)\; P(\ell_3)\\ \nn
&=& (2\pi)^6 \frac{\delta_{\bar{\ell}_1 \bar{\ell}_4}\;\delta_{\bar{\ell}_2 \bar{\ell}_5}\;\delta_{\bar{\ell}_3 \bar{\ell}_6}}{A_R(\bar{\ell}_1)\; A_R(\bar{\ell}_2)\; A_R(\bar{\ell}_3)} \int_1 \int_2 \int_3 \br{ \delta^{(2)}_{\rm D}(\vek{\ell}_1+\vek{\ell}_2+\vek{\ell}_3) }^2\; P(\ell_1)\; P(\ell_2)\; P(\ell_3)\;,
}
where the integrations over $\vek{\ell}_4$ to $\vek{\ell}_6$ only yield a non-zero result if the annuli of the angular frequencies in the corresponding delta-distributions, which are integrated over, coincide. Thus, for every such integration a Kronecker symbol is generated. The resulting expression in (\ref{eq:covcalc_2}) can now easily be simplified by using (\ref{eq:delta0}), producing a factor of $A/(2\,\pi)^2$, and subsequently (\ref{eq:gaunt2}) to execute the remaining angular integrations. Note that again only the delta-distribution depends on the polar angles of the angular frequencies. Therefore, considering only the Gaussian contribution to the covariance, (\ref{eq:cov_insertestimator}) turns into
\eqa{
\label{eq:covaftergaunt}
\ba{ \hat{B}(\bar{\ell}_1,\bar{\ell}_2,\bar{\ell}_3)\; \hat{B}(\bar{\ell}_4,\bar{\ell}_5,\bar{\ell}_6) }_{\rm Gauss} &=& \frac{(2\,\pi)^9}{A\; A_R(\bar{\ell}_1)\; A_R(\bar{\ell}_2)\; A_R(\bar{\ell}_3)}\; \Lambda^{-1} \br{\bar{\ell}_1,\bar{\ell}_2,\bar{\ell}_3}\\ \nn
&& \hspace*{1cm} \times\; D_{\bar{\ell}_1,\bar{\ell}_2,\bar{\ell}_3,\bar{\ell}_4,\bar{\ell}_5,\bar{\ell}_6}\; \int_{A_R(\bar{\ell}_1)} \frac{ \dd \ell_1 \ell_1}{A_R(\bar{\ell}_1)}\; P(\ell_1)  \int_{A_R(\bar{\ell}_2)} \frac{ \dd \ell_2 \ell_2}{A_R(\bar{\ell}_2)}\; P(\ell_2) \int_{A_R(\bar{\ell}_3)} \frac{ \dd \ell_3 \ell_3}{A_R(\bar{\ell}_3)}\; P(\ell_3)\;,
}
where we again pulled $\Lambda$, evaluated at the averaged angular frequencies, out of the radial integrations. Besides, we defined the shorthand notation
\eq{
\label{eq:kroneckers}
D_{\ell_1,\ell_2,\ell_3,\ell_4,\ell_5,\ell_6} \equiv \delta_{\ell_1 \ell_4}\;\delta_{\ell_2 \ell_5}\;\delta_{\ell_3 \ell_6} + \delta_{\ell_1 \ell_5}\;\delta_{\ell_2 \ell_4}\;\delta_{\ell_3 \ell_6} + \delta_{\ell_1 \ell_4}\;\delta_{\ell_2 \ell_6}\;\delta_{\ell_3 \ell_5} + \delta_{\ell_1 \ell_5}\;\delta_{\ell_2 \ell_6}\;\delta_{\ell_3 \ell_4} + \delta_{\ell_1 \ell_6}\;\delta_{\ell_2 \ell_4}\;\delta_{\ell_3 \ell_5} +\delta_{\ell_1 \ell_6}\;\delta_{\ell_2 \ell_5}\;\delta_{\ell_3 \ell_4}
}
for convenience. By making use of (\ref{eq:annuli}) and defining the bin-averaged power spectrum as
\eq{
\label{eq:psaverage}
P(\bar{\ell}_i) \equiv \int_{\bar{\ell}_i-1/2 \Delta \ell}^{\bar{\ell}_i+1/2 \Delta \ell} \frac{ \dd \ell_i \ell_i}{\bar{\ell}_i \Delta \ell} \; P(\ell_i)\;,
}
see \citet{joachimi08}, in analogy to the definition of the bin-averaged bispectrum, one obtains the expression
\eq{
\label{eq:bicovgauss}
\ba{ \hat{B}(\bar{\ell}_1,\bar{\ell}_2,\bar{\ell}_3)\; \hat{B}(\bar{\ell}_4,\bar{\ell}_5,\bar{\ell}_6) }_{\rm Gauss} = \frac{(2\,\pi)^3}{A\; \bar{\ell}_1 \bar{\ell}_2 \bar{\ell}_3 \Delta \ell_1 \Delta \ell_2 \Delta \ell_3 }\; \Lambda^{-1} \br{\bar{\ell}_1,\bar{\ell}_2,\bar{\ell}_3}\; D_{\bar{\ell}_1,\bar{\ell}_2,\bar{\ell}_3,\bar{\ell}_4,\bar{\ell}_5,\bar{\ell}_6}\; P(\bar{\ell}_1) P(\bar{\ell}_2) P(\bar{\ell}_3)\;.
}

Terms composed of two three-point correlators can be processed as follows,
\eqa{
\label{eq:covcalc_3}
&& \int_1 \int_2 \int_3 \int_4 \int_5 \int_6 \delta^{(2)}_{\rm D}(\vek{\ell}_1+\vek{\ell}_2+\vek{\ell}_3)\; \delta^{(2)}_{\rm D}(\vek{\ell}_4+\vek{\ell}_5+\vek{\ell}_6)\; \ba{g(\vek{\ell}_1)\; g(\vek{\ell}_2)\; g(\vek{\ell}_4)} \ba{g(\vek{\ell}_3)\; g(\vek{\ell}_5)\; g(\vek{\ell}_6)}\\ \nn
&=& (2\pi)^4 \int_1 \int_2 \int_3 \int_4 \int_5 \int_6 \delta^{(2)}_{\rm D}(\vek{\ell}_1+\vek{\ell}_2+\vek{\ell}_3)\; \delta^{(2)}_{\rm D}(\vek{\ell}_4+\vek{\ell}_5+\vek{\ell}_6)\; \delta^{(2)}_{\rm D}(\vek{\ell}_1+\vek{\ell}_2+\vek{\ell}_4)\; \delta^{(2)}_{\rm D}(\vek{\ell}_3+\vek{\ell}_5+\vek{\ell}_6)\; B(\ell_1,\ell_2,\ell_4)\; B(\ell_3,\ell_5,\ell_6)\\ \nn
&=& (2\pi)^4\; \delta_{\bar{\ell}_3 \bar{\ell}_4} \int_1 \int_2 \int_3 \int_5 \int_6 \delta^{(2)}_{\rm D}(\vek{\ell}_1+\vek{\ell}_2+\vek{\ell}_3)\; \delta^{(2)}_{\rm D}(-\vek{\ell}_1-\vek{\ell}_2+\vek{\ell}_5+\vek{\ell}_6)\; \delta^{(2)}_{\rm D}(\vek{\ell}_3+\vek{\ell}_5+\vek{\ell}_6)\; B(\ell_1,\ell_2,|\vek{\ell}_1+\vek{\ell}_2|)\; B(\ell_3,\ell_5,\ell_6)\\ \nn
&=& A\; (2\pi)^2\; \delta_{\bar{\ell}_3 \bar{\ell}_4} \int_1 \int_2 \int_3 \int_5 \int_6 \delta^{(2)}_{\rm D}(\vek{\ell}_1+\vek{\ell}_2+\vek{\ell}_3)\; \delta^{(2)}_{\rm D}(\vek{\ell}_3+\vek{\ell}_5+\vek{\ell}_6)\; B(\ell_1,\ell_2,\ell_3)\; B(\ell_3,\ell_5,\ell_6)\;,
}
where to generate the Kronecker symbol $\delta_{\bar{\ell}_3 \bar{\ell}_4}$, we made use of fact that $\vek{\ell}_1+\vek{\ell}_2=-\vek{\ell}_3$ due to the corresponding delta-distribution. To arrive at the last equality, (\ref{eq:delta0}) has been applied after processing the arguments of the delta-distributions similar to (\ref{eq:deltacomb}). The remaining terms, containing four- and six-point correlators of $g$, can be dealt with in close analogy to (\ref{eq:covcalc_3}). We mention the special case
\eqa{
\label{eq:covcalc_3special}
&& \int_1 \int_2 \int_3 \int_4 \int_5 \int_6 \delta^{(2)}_{\rm D}(\vek{\ell}_1+\vek{\ell}_2+\vek{\ell}_3)\; \delta^{(2)}_{\rm D}(\vek{\ell}_4+\vek{\ell}_5+\vek{\ell}_6)\; \ba{g(\vek{\ell}_1)\; g(\vek{\ell}_2)\; g(\vek{\ell}_3)} \ba{g(\vek{\ell}_4)\; g(\vek{\ell}_5)\; g(\vek{\ell}_6)}\\ \nn
&=& (2\pi)^4 \int_1 \int_2 \int_3 \int_4 \int_5 \int_6 \br{ \delta^{(2)}_{\rm D}(\vek{\ell}_1+\vek{\ell}_2+\vek{\ell}_3) }^2\; \br{ \delta^{(2)}_{\rm D}(\vek{\ell}_4+\vek{\ell}_5+\vek{\ell}_6) }^2 B(\ell_1,\ell_2,\ell_3)\; B(\ell_4,\ell_5,\ell_6)\\ \nn
&=& A^2 \int_1 \int_2 \int_3 \delta^{(2)}_{\rm D}(\vek{\ell}_1+\vek{\ell}_2+\vek{\ell}_3)\; B(\ell_1,\ell_2,\ell_3) \int_4 \int_5 \int_6 \delta^{(2)}_{\rm D}(\vek{\ell}_4+\vek{\ell}_5+\vek{\ell}_6)\; B(\ell_4,\ell_5,\ell_6)\\ \nn
&=& \frac{A^2}{(2\pi)^4}\; \Lambda \br{\bar{\ell}_1,\bar{\ell}_2,\bar{\ell}_3}\; \Lambda \br{\bar{\ell}_4,\bar{\ell}_5,\bar{\ell}_6}\; B(\bar{\ell}_1,\bar{\ell}_2,\bar{\ell}_3)\; B(\bar{\ell}_4,\bar{\ell}_5,\bar{\ell}_6)\;,
}
which, after inserting this expression into (\ref{eq:cov_insertestimator}), cancels the product $B(\bar{\ell}_1,\bar{\ell}_2,\bar{\ell}_3) B(\bar{\ell}_4,\bar{\ell}_5,\bar{\ell}_6)$ in the definition of the covariance (\ref{eq:defcov}).

Combining these results, we obtain the total bispectrum covariance
\eqa{
\label{eq:bicovtot}
{\rm Cov} \br{B(\bar{\ell}_1,\bar{\ell}_2,\bar{\ell}_3),\; B(\bar{\ell}_4,\bar{\ell}_5,\bar{\ell}_6)} &=& \frac{(2\,\pi)^3}{A\; \bar{\ell}_1 \bar{\ell}_2 \bar{\ell}_3 \Delta \ell_1 \Delta \ell_2 \Delta \ell_3 }\; \Lambda^{-1} \br{\bar{\ell}_1,\bar{\ell}_2,\bar{\ell}_3}\; D_{\bar{\ell}_1,\bar{\ell}_2,\bar{\ell}_3,\bar{\ell}_4,\bar{\ell}_5,\bar{\ell}_6}\; P(\bar{\ell}_1) P(\bar{\ell}_2) P(\bar{\ell}_3)\\ \nn
&& \hspace*{-3cm} + \frac{\cal C}{A}\; \delta_{\bar{\ell}_3 \bar{\ell}_4} \int_1 \int_2 \int_3 \int_5 \int_6 \delta^{(2)}_{\rm D}(\vek{\ell}_1+\vek{\ell}_2+\vek{\ell}_3)\; \delta^{(2)}_{\rm D}(\vek{\ell}_3+\vek{\ell}_5+\vek{\ell}_6)\; B(\ell_1,\ell_2,\ell_3)\; B(\ell_3,\ell_5,\ell_6) + (\mbox{8 perm.})\\ \nn
&& \hspace*{-3cm} + \frac{\cal C}{A}\; \delta_{\bar{\ell}_3 \bar{\ell}_6} \int_1 \int_2 \int_3 \int_4 \int_5 \delta^{(2)}_{\rm D}(\vek{\ell}_1+\vek{\ell}_2+\vek{\ell}_3)\; \delta^{(2)}_{\rm D}(\vek{\ell}_4+\vek{\ell}_5-\vek{\ell}_3)\; P_4(\vek{\ell}_1,\vek{\ell}_2,\vek{\ell}_4,\vek{\ell}_5)\; P(\ell_3) + (\mbox{8 perm.})\\ \nn
&& \hspace*{-3cm} + \frac{\cal C}{A} \int_1 \int_2 \int_3 \int_4 \int_5 \int_6 \delta^{(2)}_{\rm D}(\vek{\ell}_1+\vek{\ell}_2+\vek{\ell}_3)\; \delta^{(2)}_{\rm D}(\vek{\ell}_4+\vek{\ell}_5+\vek{\ell}_6)\; P_6(\vek{\ell}_1,\vek{\ell}_2,\vek{\ell}_3,\vek{\ell}_4,\vek{\ell}_5,\vek{\ell}_6)\;,
}
where the prefactor reads ${\cal C} \equiv (2\pi)^6\; \Lambda^{-1} \br{\bar{\ell}_1,\bar{\ell}_2,\bar{\ell}_3}\; \Lambda^{-1} \br{\bar{\ell}_4,\bar{\ell}_5,\bar{\ell}_6}$.

The general form of the covariance terms is in agreement with the expressions derived in \citet{sefusatti06}. As mentioned in Sect.$\,$\ref{sec:estimator}, shot or shape noise can readily be included into this covariance by adding a corresponding noise term to the power spectra. Weak lensing or galaxy clustering surveys often have in addition tomographic information, so that the data is binned into (photometric) redshift bins. The covariance can be generalized to this case in a straightforward manner by obeying the practical rule that each photometric redshift \lq sticks\rq\ to the angular frequency it is assigned to, see \citet{takada04}. A similar argument holds for the generalization to CMB polarization bispectrum covariances \citep{hu00}.

\section{Equivalence to spherical harmonics approach}
\label{sec:equivalence}

In this section we demonstrate that both our and the spherical harmonic approach are equivalent in the sense that they measure the same information in a survey. Moreover, we investigate the behavior with respect to parity, and the relation between the covariances of both approaches, considering for the remainder of this work only the Gaussian part of (\ref{eq:bicovtot}).

\subsection{Comparison of covariances}
\label{sec:comp}

On the celestial sphere one can decompose the random field $g$ into spherical harmonics, which produces a set of coefficients $g_{\ell m}$ with $\ell,m$ integers and $\ell \geq 0$, $-\ell \leq m \leq \ell$. In terms of the $g_{\ell m}$ one can define a bispectrum estimator as \citep[e.g.][]{hu00}
\eq{
\label{eq:fullskyestimator}
\hat{B}_{\ell_1, \ell_2, \ell_3} = \sum_{m_1,m_2,m_3} \wigner{\ell_1}{\ell_2}{\ell_3}{m_1}{m_2}{m_3}\; g_{\ell_1 m_1}\; g_{\ell_2 m_2}\; g_{\ell_3 m_3}\;,
}
where the object in parentheses is the Wigner-$3j$ symbol. Properties of the Wigner symbol are reviewed in \citet{hu00}; most importantly, it obeys the triangle condition, i.e. it is non-zero only for $|\ell_1 - \ell_2| \leq \ell_3 \leq \ell_1 + \ell_2$ and permutations thereof. For this estimator \citet{hu00} derived the simple Gaussian covariance
\eq{
\label{eq:fullskycov}
{\rm Cov} \br{B_{\bar{\ell}_1,\bar{\ell}_2,\bar{\ell}_3},\; B_{\bar{\ell}_4,\bar{\ell}_5,\bar{\ell}_6}} = D_{\bar{\ell}_1,\bar{\ell}_2,\bar{\ell}_3,\bar{\ell}_4,\bar{\ell}_5,\bar{\ell}_6}\; P_{\bar{\ell}_1} P_{\bar{\ell}_2} P_{\bar{\ell}_3}\;,
}
where $P_{\ell}$ denotes the full-sky power spectrum, and where $D_{\bar{\ell}_1,\bar{\ell}_2,\bar{\ell}_3,\bar{\ell}_4,\bar{\ell}_5,\bar{\ell}_6}$ is used as defined in (\ref{eq:kroneckers}). Moreover, he gives approximate relations between the spherical harmonic and Fourier-plane spectra,
\eq{
\label{eq:spectrarelations}
P_{\ell} \approx P(\ell)\;; \hspace*{1cm}
B_{\ell_1, \ell_2, \ell_3} \approx \wigner{\ell_1}{\ell_2}{\ell_3}{0}{0}{0}\; \sqrt{ \frac{(2\ell_1+1)(2\ell_2+1)(2\ell_3+1)}{4\pi} }\; B(\ell_1, \ell_2, \ell_3)\;,
}
valid for $\ell_1, \ell_2, \ell_3 \gg 1$. These equations can only hold for integer $\ell$. In addition, the Wigner symbol with $m_1=m_2=m_3=0$ vanishes for $L \equiv \ell_1+\ell_2+\ell_3$ odd, see the following section for details. Making use of the standard procedure of multiplying (\ref{eq:fullskycov}) by an \emph{ad hoc} factor of $f^{-1}_{\rm sky}=4\pi/A$ to account for finite sky coverage of the survey, one can derive a flat-sky spherical harmonic covariance with (\ref{eq:spectrarelations}) as \citep{hu00,takada04}
\eq{
\label{eq:bicovtj}
\ba{ \hat{B}(\bar{\ell}_1,\bar{\ell}_2,\bar{\ell}_3)\; \hat{B}(\bar{\ell}_4,\bar{\ell}_5,\bar{\ell}_6) } \approx \frac{(4\pi)^2\; D_{\bar{\ell}_1,\bar{\ell}_2,\bar{\ell}_3,\bar{\ell}_4,\bar{\ell}_5,\bar{\ell}_6}}{A\;(2 \bar{\ell}_1 + 1)\,(2 \bar{\ell}_2 + 1)\,(2 \bar{\ell}_3 + 1)} \; \wigner{\bar{\ell}_1}{\bar{\ell}_2}{\bar{\ell}_3}{0}{0}{0}^{-2} P(\bar{\ell}_1)\; P(\bar{\ell}_2)\; P(\bar{\ell}_3)\;,
}
where still the angular frequencies are required to be integer, and $L$ even. As is true for our approach, (\ref{eq:bicovtj}) holds for $\ell \gg 1$ only. To be able to compare this widely used formula to our results, a relation between the Wigner symbol and $\Lambda$ has to be found.

When comparing the spherical harmonics and the Fourier-plane approach, \citet{hu00} already came across integrals of the form (\ref{eq:gaunt2}). We reproduce his computation,
\eqa{
\label{eq:gaunt}
&& \int \dd^2 \ell_1 \int \dd^2 \ell_2 \int \dd^2 \ell_3\; \delta^{(2)}_{\rm D}(\vek{\ell}_1+\vek{\ell}_2+\vek{\ell}_3) \; = \; \int \dd^2 \ell_1 \int \dd^2 \ell_2 \int \dd^2 \ell_3\; \int \frac{\dd^2 \theta}{(2\,\pi)^2} \expo{\ic \br{\svek{\ell}_1 + \svek{\ell}_2 + \svek{\ell}_3} \cdot \svek{\theta}}\\ \nn
&\approx& \int \dd \ell_1 \ell_1 \int \dd \ell_2 \ell_2 \int \dd \ell_3 \ell_3 \sqrt{\frac{(2\,\pi)^5}{\ell_1 \ell_2 \ell_3}} \int \dd \Omega\; Y_{\ell_1}^0(\vek{n})\; Y_{\ell_2}^0(\vek{n})\; Y_{\ell_3}^0(\vek{n}) \; \approx \; 8 \pi^2 \int \dd \ell_1 \ell_1 \int \dd \ell_2 \ell_2 \int \dd \ell_3 \ell_3 \wigner{\ell_1}{\ell_2}{\ell_3}{0}{0}{0}^{2}\;,
}
where $\int \dd \Omega$ is the integral over the unit sphere, and where $Y_\ell^m(\vek{n})$ denotes the spherical harmonic function with $\vek{n}$ the unit normal vector on the sphere. We are concerned with the validity of this derivation for the following reasons: Terms with integer and real-valued $\ell$ are mixed, e.g. it remains unclear how the integration over the Wigner symbol squared is to be understood. To get from the second to the third equality, the Fourier base $\expo{\ic \svek{\ell} \cdot \svek{\theta}}$ is expanded into spherical harmonics, an approximation which \citet{hu00} correctly states to be valid for small angles only. However, the integration over angles runs over the full two-dimensional plane or the unit sphere, respectively. Moreover, it is not specified how the non-trivial transition from an integral over the plane to one over the unit sphere is executed. Instead of (\ref{eq:gaunt}), we propose to use (\ref{eq:gaunt2}), which is an exact and rigorous expression.

To allow for a comparison between (\ref{eq:gaunt}) and our approach based on (\ref{eq:gaunt2}), we need to establish a relation between the square of the Wigner symbol and (\ref{eq:deflambda}). We refer to \citet[see also references therein]{borodin78} who compute approximation formulae of the Wigner symbol in the context of the quasi-continuous limit of quantum states with high angular momenta. The base of their derivation is formed by the exact relation
\eq{
\label{eq:wignerD}
\int^{2\pi}_0 \dd \varphi \int^{2\pi}_0 \dd \psi \int^{\pi}_0 \dd \theta \sin \theta\; D^{\ell_1}_{m_1 m'_1}\br{\varphi,\theta,\psi}\; D^{\ell_2}_{m_2 m'_2}\br{\varphi,\theta,\psi}\; D^{\ell_3}_{m_3 m'_3}\br{\varphi,\theta,\psi} = 8 \pi^2\; \wigner{\ell_1}{\ell_2}{\ell_3}{m_1}{m_2}{m_3} \cdot \wigner{\ell_1}{\ell_2}{\ell_3}{m'_1}{m'_2}{m'_3}\;,
}
where $D^{\ell}_{m m'}$ denotes the $m \times m'$ element of the Wigner D matrix, which in turn is a function of the three Euler angles $\varphi$, $\theta$, and $\psi$. Making use of a quasi-classical approximation of the $D^{\ell}_{m m'}$, \citet{borodin78} compute expressions for the general Wigner symbol in the limit of large and continuous angular frequencies. From these results we extract the approximation
\eqa{
\label{eq:wignerproxy}
\wigner{\ell_1}{\ell_2}{\ell_3}{0}{0}{0}^2 &\approx& \frac{2}{\pi}\; \Biggl\{ 2\br{\ell_1+\frac{1}{2}}^2\; \br{\ell_2+\frac{1}{2}}^2 + 2\br{\ell_2+\frac{1}{2}}^2\; \br{\ell_3+\frac{1}{2}}^2\\ \nn
 && \hspace*{5cm} +\; 2\br{\ell_3+\frac{1}{2}}^2\; \br{\ell_1+\frac{1}{2}}^2 - \br{\ell_1+\frac{1}{2}}^4 - \br{\ell_2+\frac{1}{2}}^4 - \br{\ell_3+\frac{1}{2}}^4 \Biggr\}^{-1/2}\;,
}
which allows us to generalize the Wigner symbol to real-valued arguments. Equation (\ref{eq:wignerproxy}) holds only for $\ell_1, \ell_2, \ell_3 \gg 1$, which, in the quantum-mechanical context of \citet{borodin78}, originates from the use of expressions that are valid for large angular momenta, i.e. the quasi-classical limit, only. This condition on angular frequencies also underlies the approximations in (\ref{eq:spectrarelations}) and (\ref{eq:gaunt}) and can in our context be interpreted as a natural consequence of working in the flat-sky approximation.

\begin{figure}[t]
\centering
\includegraphics[scale=.65,angle=270]{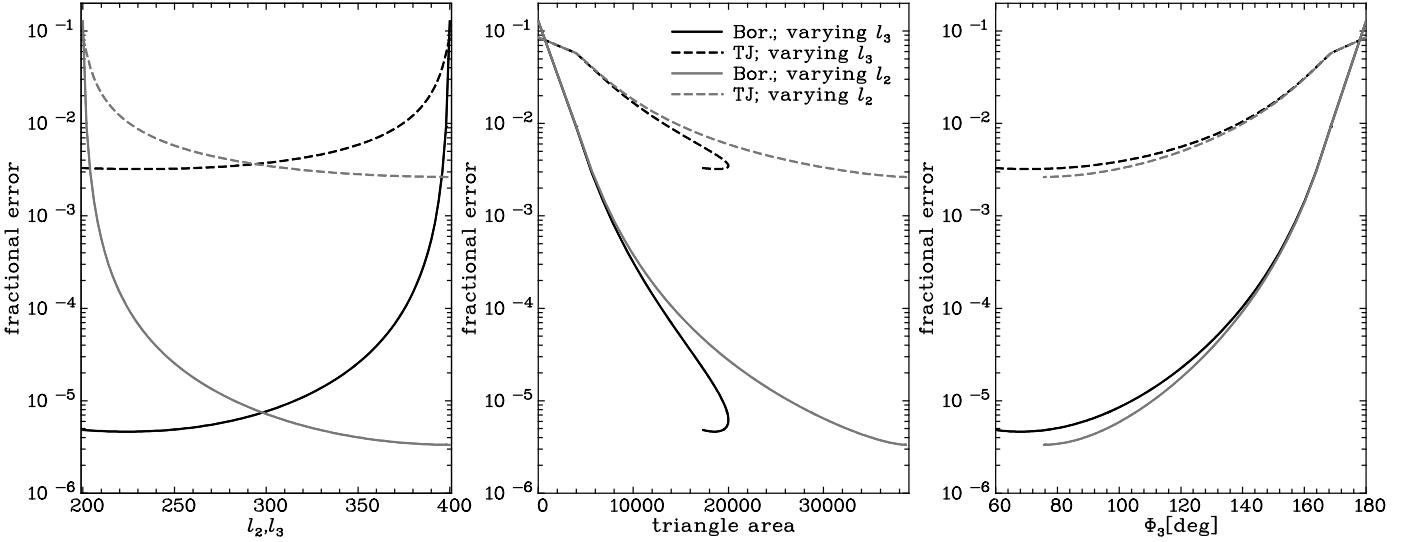}
\caption{
Fractional error of the approximation formulae for the Wigner symbol. \textit{Left panel}: Shown are the relative deviations of (\ref{eq:wignerproxy}) and (\ref{eq:wignerproxyTJ04}) from the true absolute value of the Wigner symbol. The same triangle configurations as in Fig.$\,$\ref{fig:triangleareas} are used. Results for $\ell_1=200,\ell_3=400$ and varying $\ell_2$ are shown in gray while those corresponding to $\ell_1,\ell_2=200$ and varying $\ell_3$ are plotted in black. Solid curves are obtained using (\ref{eq:wignerproxy}), dashed curves by employing (\ref{eq:wignerproxyTJ04}). \textit{Center panel}: Same as above, but now plotting on the abscissa the corresponding triangle area enclosed by the three angular frequency vectors. Note that before reaching the equilateral configuration, the area has a maximum and starts to decrease again. \textit{Right panel}: Same as above, but now as a function of the internal angle $\Phi_3$ opposite $\ell_3$, which is the longest side of the triangle in both configurations considered. Hence, $\Phi_3=60^\circ$ corresponds to the equilateral case, and $\Phi_3=180^\circ$ to the degenerate case.
}
\label{fig:wigner_compare}
\end{figure}

As is demonstrated in Fig.$\,$\ref{fig:wigner_compare}, we find that (\ref{eq:wignerproxy}) constitutes an excellent approximation, whose accuracy over a wide range of $\ell$-values is orders of magnitude better than the approximation given in \citet{takada04}, Eq.$\,$(A3),
\eqa{
\label{eq:wignerproxyTJ04}
\wigner{\ell_1}{\ell_2}{\ell_3}{0}{0}{0}^2 &\approx& \frac{ \expo{3}}{\sqrt{2}\, \pi}\; (L+2)^{-1/2}\; \br{\frac{L}{2}-\ell_1+1}^{-1/2}\; \br{\frac{L}{2}-\ell_2+1}^{-1/2}\; \br{\frac{L}{2}-\ell_3+1}^{-1/2}\;  \\ \nn
 && \hspace*{3cm} \times \; \br{\frac{L/2-\ell_1+1/2}{L/2-\ell_1+1}}^{L-2\ell_1+1/2}\; \br{\frac{L/2-\ell_2+1/2}{L/2-\ell_2+1}}^{L-2\ell_2+1/2}\; \br{\frac{L/2-\ell_3+1/2}{L/2-\ell_3+1}}^{L-2\ell_3+1/2}\;.
}
Only for triangle configurations close to degeneracy does the latter formula perform slightly better. Both approximation formulae are least accurate in the case of a degenerate triangle configuration with fractional errors around $10\,\%$ or slightly above, but improve quickly to very small percentage deviations when the configuration approaches a more equilateral form. In Fig.$\,$\ref{fig:wigner_compare} we also plot the fractional errors as a function of the triangle area enclosed by the three angular frequency vectors and as a function of the internal angle $\Phi_3$ opposite $\ell_3$, being the longest side of the triangle in the configurations considered. In terms of these quantities we observe a more universal behavior of the errors, in particular in the regime where the approximations are less accurate. We find to good approximation that, when approaching the degenerate case, relative errors increase exponentially with decreasing triangle area and increasing $\Phi_3$.

For $\ell \gg 1$, and if the triangle configuration is not too close to the degenerate case, one may approximate $\ell_i + 1/2 \approx \ell_i$, so that one finds from (\ref{eq:deflambda}) and (\ref{eq:wignerproxy})
\eq{
\label{eq:wignerlambda}
\wigner{\ell_1}{\ell_2}{\ell_3}{0}{0}{0}^2 \approx \frac{\Lambda \br{\ell_1,\ell_2,\ell_3}}{2\pi}\;.
}
Remarkably, since for integer angular frequencies we have $\Delta \ell_1 = \Delta \ell_2 = \Delta \ell_3 =1$, (\ref{eq:wignerproxy}) exactly reproduces our earlier conjecture (\ref{eq:areaproxy2}), which strongly supports its validity. If one replaces the Wigner symbol in (\ref{eq:gaunt}) by (\ref{eq:wignerlambda}), however, one obtains a result which is a factor of 2 larger compared to (\ref{eq:gaunt2}).

Inserting (\ref{eq:wignerlambda}) into (\ref{eq:bicovtj}), and using $2\ell+1 \approx 2\ell$ for $\ell \gg 1$, we get
\eq{
\label{eq:bicovtj2}
\ba{ \hat{B}(\bar{\ell}_1,\bar{\ell}_2,\bar{\ell}_3)\; \hat{B}(\bar{\ell}_4,\bar{\ell}_5,\bar{\ell}_6) } \approx \frac{2\pi^2\; D_{\bar{\ell}_1,\bar{\ell}_2,\bar{\ell}_3,\bar{\ell}_4,\bar{\ell}_5,\bar{\ell}_6}}{A\; \bar{\ell}_1\, \bar{\ell}_2\, \bar{\ell}_3 } \; \wigner{\bar{\ell}_1}{\bar{\ell}_2}{\bar{\ell}_3}{0}{0}{0}^{-2} P(\bar{\ell}_1)\; P(\bar{\ell}_2)\; P(\bar{\ell}_3)\;,
}
which is equivalent to (\ref{eq:bicovgauss}) if the latter equation is specified to $\Delta \ell_1 = \Delta \ell_2 = \Delta \ell_3 =1$, and integer $\ell$ with $L$ even -- except for (\ref{eq:bicovtj2}) being a factor of 2 smaller. In the following, we are going to elaborate on this apparent discrepancy.

\subsection{Parity}
\label{sec:parity}

To elucidate the different noise properties of the Fourier-plane and spherical harmonic bispectrum estimators, we investigate their behavior with respect to parity. In two dimensions the parity transformation corresponds to an axis reflection, or equivalently, the reversal of the polar angle of all spatial vectors. To flip the parity of a triangle, one can do an odd permutation of its sides, see e.g. the two triangles sketched in Fig.$\,$\ref{fig:sketch_overlap}. Hence, to test the behavior of estimators for triangles of different parity, it is sufficient to flip any two of its angular frequency arguments.

Consulting (\ref{eq:fullskyestimator}), we find
\eq{
\label{eq:fullskyparity}
\hat{B}_{\ell_1, \ell_3, \ell_2} = (-1)^{\ell_1+\ell_2+\ell_3}\; \hat{B}_{\ell_1, \ell_2, \ell_3}
}
because of the behavior of the Wigner symbol under change of parity,
\eq{
\label{eq:paritywigner}
\wigner{\ell_1}{\ell_2}{\ell_3}{m_1}{m_2}{m_3} = (-1)^{\ell_1+\ell_2+\ell_3}\; \wigner{\ell_1}{\ell_3}{\ell_2}{m_1}{m_3}{m_2}\;,
}
and likewise for all odd permutations of the columns in the Wigner symbol. Thus, the spherical harmonics estimator is parity-invariant for $L$ even and changes sign for $L$ odd. Most cosmological theories predict parity-invariant large-scale structures and CMB anisotropies. If parity symmetry is built into the cosmological model at consideration, measures that vary under parity transformations do not have any predictive power, wherefore they are usually not considered in a data analysis. Accordingly, (\ref{eq:fullskyestimator}) is only used for arguments that have $L$ even. Note that parity invariance is also incorporated into the relation between the spherical harmonics and Fourier-plane bispectra, see the second equality of (\ref{eq:spectrarelations}), via the Wigner symbol which vanishes for $L$ odd (this behavior is a direct consequence of (\ref{eq:paritywigner}) for $m_1=m_2=m_3=0$).

The Fourier-plane estimator is by design parity-invariant, which can be seen mathematically from swapping arguments of (\ref{eq:estimator}), or illustratively by inspecting Fig.$\,$\ref{fig:sketch_overlap}. From the sketch it is evident that triangle configurations of different parity are averaged over with equal weight. For a more formal argument, we can explicitly construct estimators that average only over triangle configurations of the same parity. To this end, consider the two-dimensional cross product $\vek{a} \times \vek{b} = a_x b_y - a_y b_x$ \citep{schneider03b} of the angular frequency vectors $\vek{\ell}_1,\vek{\ell}_2,\vek{\ell}_3$. If they form a triangle, one finds $\vek{\ell}_1 \times \vek{\ell}_2 = \vek{\ell}_2 \times \vek{\ell}_3 = \vek{\ell}_3 \times \vek{\ell}_1$, which follows from $\vek{\ell}_1+\vek{\ell}_2+\vek{\ell}_3=0$. A change in the parity of the triangle implies a sign flip in these cross products. 

Noting that $\vek{\ell}_i \times \vek{\ell}_j = \ell_i \ell_j\, \sin (\varphi_{\ell_j} - \varphi_{\ell_i})$, we compute a condition on the polar angles,
\eq{
\label{eq:condfortriangle}
\varphi_{\ell_2} - \varphi_{\ell_1} \in \bb{0,\pi}\;; ~~~ \varphi_{\ell_3} - \varphi_{\ell_2} \in \bb{0,\pi}\;; ~~~ \varphi_{\ell_1} - \varphi_{\ell_3} \in \bb{0,\pi}\;.
}
To obtain the parity transformed triangle, swap the signs of the polar angles in (\ref{eq:condfortriangle}). Under the premise that the vectors do form a triangle, one of the conditions in (\ref{eq:condfortriangle}) is redundant, the remaining ones restricting the angular integrations in the averaging of (\ref{eq:estimator}). For instance, the integration ranges could be modified to $\varphi_{\ell_1} \in \bb{0,2\pi}$, $\varphi_{\ell_2} \in \bb{\varphi_{\ell_1},\pi+\varphi_{\ell_1}}$, and $\varphi_{\ell_3} \in \bb{\varphi_{\ell_1}-\pi,\varphi_{\ell_1}}$. Due to rotational symmetry, which still holds, the inner integrals have to yield the same result for all possible values of $\varphi_{\ell_1}$. Therefore, we can set the ranges of the inner integrals to $\varphi_{\ell_2} \in \bb{0,\pi}$ and $\varphi_{\ell_3} \in \bb{-\pi,0}$ without loss of generality. To maintain the symmetry, we keep the integral over $\varphi_{\ell_1}$ in our notation. These findings are reflected in the shorthand notation
\eq{
\label{eq:parityint}
\int \dd\!\bc{\varphi_1,\varphi_2,\varphi_3} \equiv \int_0^{2\pi} \frac{\dd \varphi_{\ell_1}}{2\pi} \int_0^{\pi} \frac{\dd \varphi_{\ell_2}}{\pi} \int_{-\pi}^0 \frac{\dd \varphi_{\ell_3}}{\pi}\;,
}
which we use to define the following bispectrum estimators,
\eqa{
\label{eq:estimatorparity}
\hat{B}_\Delta(\bar{\ell}_1,\bar{\ell}_2,\bar{\ell}_3) &=& \frac{2\pi^2}{A} \Lambda^{-1} \br{\bar{\ell}_1,\bar{\ell}_2,\bar{\ell}_3} \int_{\bar{\ell}_1-1/2 \Delta \ell}^{\bar{\ell}_1+1/2 \Delta \ell} \frac{ \dd \ell_1 \ell_1}{\bar{\ell}_1 \Delta \ell_1} \int_{\bar{\ell}_2-1/2 \Delta \ell}^{\bar{\ell}_2+1/2 \Delta \ell} \frac{ \dd \ell_2 \ell_2}{\bar{\ell}_2 \Delta \ell_2} \int_{\bar{\ell}_3-1/2 \Delta \ell}^{\bar{\ell}_3+1/2 \Delta \ell} \frac{ \dd \ell_3 \ell_3}{\bar{\ell}_3 \Delta \ell_3}\\ \nn
&& \hspace*{1cm} \times\; \frac{1}{3} \bc{ \int \dd\!\bc{\varphi_1,\varphi_2,\varphi_3} + \int \dd\!\bc{\varphi_2,\varphi_3,\varphi_1} + \int \dd\!\bc{\varphi_3,\varphi_1,\varphi_2} }\; \delta^{(2)}_{\rm D}(\vek{\ell}_1+\vek{\ell}_2+\vek{\ell}_3)\; g(\vek{\ell}_1)\; g(\vek{\ell}_2)\; g(\vek{\ell}_3)\;;\\ \nn
\hat{B}_\nabla(\bar{\ell}_1,\bar{\ell}_2,\bar{\ell}_3) &=& \frac{2\pi^2}{A} \Lambda^{-1} \br{\bar{\ell}_1,\bar{\ell}_2,\bar{\ell}_3} \int_{\bar{\ell}_1-1/2 \Delta \ell}^{\bar{\ell}_1+1/2 \Delta \ell} \frac{ \dd \ell_1 \ell_1}{\bar{\ell}_1 \Delta \ell_1} \int_{\bar{\ell}_2-1/2 \Delta \ell}^{\bar{\ell}_2+1/2 \Delta \ell} \frac{ \dd \ell_2 \ell_2}{\bar{\ell}_2 \Delta \ell_2} \int_{\bar{\ell}_3-1/2 \Delta \ell}^{\bar{\ell}_3+1/2 \Delta \ell} \frac{ \dd \ell_3 \ell_3}{\bar{\ell}_3 \Delta \ell_3}\\ \nn
&& \hspace*{1cm} \times\;  \frac{1}{3} \bc{ \int \dd\!\bc{\varphi_1,\varphi_3,\varphi_2} + \int \dd\!\bc{\varphi_2,\varphi_1,\varphi_3} + \int \dd\!\bc{\varphi_3,\varphi_2,\varphi_1} }\; \delta^{(2)}_{\rm D}(\vek{\ell}_1+\vek{\ell}_2+\vek{\ell}_3)\; g(\vek{\ell}_1)\; g(\vek{\ell}_2)\; g(\vek{\ell}_3)\;.
}
Here, we have symmetrized the restricted integrations (\ref{eq:parityint}) by averaging over all either even or odd permutations of $\bc{\varphi_{\ell_1},\varphi_{\ell_2},\varphi_{\ell_3}}$. Consequently, changing parity via any odd permutation of the angular frequencies in the arguments of (\ref{eq:estimatorparity}) turns one estimator into the other, as demanded, for instance $\hat{B}_\Delta(\bar{\ell}_1,\bar{\ell}_3,\bar{\ell}_2) = \hat{B}_\nabla(\bar{\ell}_1,\bar{\ell}_2,\bar{\ell}_3)$. 

Note that the prefactor of the estimators in (\ref{eq:estimatorparity}) is diminished by a factor of 2 with respect to (\ref{eq:estimator}), which is necessary to keep them unbiased. This can be shown by computing the expectation value of (\ref{eq:estimatorparity}) in close analogy to the procedure outlined in Sect.$\,$\ref{sec:estimator}. However, the separate consideration of angular and radial integrals that enabled us to make use of (\ref{eq:gaunt2}) is not possible anymore in this non-symmetric case. For instance, given fixed $\vek{\ell}_1$, the restricted angular integrations (\ref{eq:parityint}) can still produce a triangle of opposite parity by including a triangle with $|\vek{\ell}'_2|=\ell_3$ and $|\vek{\ell}'_3|=\ell_2$. This is reflected in the fact that the integration (\ref{eq:gaunt2}), if properly normalized\footnote{In the derivation of Sect.$\,$\ref{sec:estimator} the proper normalization of $2\pi$ for each angular integral is hidden within $A_R(\ell)$. Note that we have given (\ref{eq:gaunt2}) without this normalization, whereas it is included in (\ref{eq:geometriccorrespondence}).}, still yields the same result when limiting the length of the integration range to $\pi$.

Instead, one can execute the integral over the angular frequency which is still averaged over the full two-dimensional plane, such as
\eqa{
\label{eq:paritynewaverage1}
&& \int_{\bar{\ell}_1-1/2 \Delta \ell}^{\bar{\ell}_1+1/2 \Delta \ell} \frac{ \dd \ell_1 \ell_1}{\bar{\ell}_1 \Delta \ell_1} \int_{\bar{\ell}_2-1/2 \Delta \ell}^{\bar{\ell}_2+1/2 \Delta \ell} \frac{ \dd \ell_2 \ell_2}{\bar{\ell}_2 \Delta \ell_2} \int_{\bar{\ell}_3-1/2 \Delta \ell}^{\bar{\ell}_3+1/2 \Delta \ell} \frac{ \dd \ell_3 \ell_3}{\bar{\ell}_3 \Delta \ell_3} \int_0^{2\pi} \frac{\dd \varphi_{\ell_1}}{2\pi} \int_0^{\pi} \frac{\dd \varphi_{\ell_2}}{\pi} \int_{-\pi}^0 \frac{\dd \varphi_{\ell_3}}{\pi}\; \delta^{(2)}_{\rm D}(\vek{\ell}_1+\vek{\ell}_2+\vek{\ell}_3)\\ \nn
&=&  \frac{1}{2\pi \bar{\ell}_1 \Delta \ell_1} \int_{\bar{\ell}_2-1/2 \Delta \ell}^{\bar{\ell}_2+1/2 \Delta \ell} \frac{ \dd \ell_2 \ell_2}{\bar{\ell}_2 \Delta \ell_2} \int_{\bar{\ell}_3-1/2 \Delta \ell}^{\bar{\ell}_3+1/2 \Delta \ell} \frac{ \dd \ell_3 \ell_3}{\bar{\ell}_3 \Delta \ell_3} \int_0^{\pi} \frac{\dd \varphi_{\ell_2}}{\pi} \int_{-\pi}^0 \frac{\dd \varphi_{\ell_3}}{\pi}\; {\mathbf 1}_{\vek{\ell}_1,\vek{\ell}_2,\vek{\ell}_3}\;,
}
where ${\mathbf 1}_{\vek{\ell}_1,\vek{\ell}_2,\vek{\ell}_3}=1$ if $\vek{\ell}_1,\vek{\ell}_2,\vek{\ell}_3$ form a triangle, and 0 else. The remaining integrations reproduce the overlapping region of the annuli for $\vek{\ell}_2$ and $\vek{\ell}_3$, as depicted in Fig.$\,$\ref{fig:sketch_overlap}. By limiting the integration to the half plane to one side of an axis collinear to $\vek{\ell}_1$, the overlap is obviously halved. Since the area of the annuli for $\vek{\ell}_2$ and $\vek{\ell}_3$ is also reduced by half each, the value of the integration should double, see (\ref{eq:geometriccorrespondence}). Following the geometrical interpretation once again, we thus arrive at
\eqa{
\label{eq:paritynewaverage2}
&& \int_{\bar{\ell}_1-1/2 \Delta \ell}^{\bar{\ell}_1+1/2 \Delta \ell} \frac{ \dd \ell_1 \ell_1}{\bar{\ell}_1 \Delta \ell_1} \int_{\bar{\ell}_2-1/2 \Delta \ell}^{\bar{\ell}_2+1/2 \Delta \ell} \frac{ \dd \ell_2 \ell_2}{\bar{\ell}_2 \Delta \ell_2} \int_{\bar{\ell}_3-1/2 \Delta \ell}^{\bar{\ell}_3+1/2 \Delta \ell} \frac{ \dd \ell_3 \ell_3}{\bar{\ell}_3 \Delta \ell_3} \int_0^{2\pi} \frac{\dd \varphi_{\ell_1}}{2\pi} \int_0^{\pi} \frac{\dd \varphi_{\ell_2}}{\pi} \int_{-\pi}^0 \frac{\dd \varphi_{\ell_3}}{\pi}\; \delta^{(2)}_{\rm D}(\vek{\ell}_1+\vek{\ell}_2+\vek{\ell}_3)\;B(\ell_1,\ell_2,\ell_3)\\ \nn
&\approx& \frac{1}{2\pi^2}\;\Lambda(\bar{\ell}_1,\bar{\ell}_2,\bar{\ell}_3)\; B(\bar{\ell}_1,\bar{\ell}_2,\bar{\ell}_3)\;.
}
Comparing this result to (\ref{eq:ensembleaverage}), the estimators (\ref{eq:estimatorparity}) have indeed to be smaller by a factor of 2 to still be unbiased.

To obtain bispectrum estimators that are completely analogous to (\ref{eq:fullskyestimator}), we define
\eq{
\label{eq:finalparityestimators}
\hat{B}_\pm(\bar{\ell}_1,\bar{\ell}_2,\bar{\ell}_3) \equiv \frac{1}{2}\; \br{\hat{B}_\Delta(\bar{\ell}_1,\bar{\ell}_2,\bar{\ell}_3) \pm \hat{B}_\nabla(\bar{\ell}_1,\bar{\ell}_2,\bar{\ell}_3)}\;.
}
As $\ba{\hat{B}_-(\bar{\ell}_1,\bar{\ell}_2,\bar{\ell}_3)}=0$ for a parity symmetric random field $g$, and $\hat{B}_-(\bar{\ell}_1,\bar{\ell}_3,\bar{\ell}_2)=-\hat{B}_-(\bar{\ell}_1,\bar{\ell}_2,\bar{\ell}_3)$, this estimator shows identical behavior compared to $\hat{B}_{\bar{\ell}_1,\bar{\ell}_2,\bar{\ell}_3}$ with $L$ odd. In practice both measures could be used to assess deviations from parity symmetry. The estimators $\hat{B}_+(\bar{\ell}_1,\bar{\ell}_2,\bar{\ell}_3)$ and $\hat{B}_{\bar{\ell}_1,\bar{\ell}_2,\bar{\ell}_3}$ with $L$ even are likewise invariant under parity transformations. After some algebra that closely follows the outline of Sect.$\,$\ref{sec:covariance} we find that the covariance of $\hat{B}_+$ is the same as (\ref{eq:bicovtot}), which is not unexpected because we already noted that (\ref{eq:estimator}) is also parity-symmetric.

With (\ref{eq:finalparityestimators}) at hand, one can readily extract the different treatment of even and odd parity measures in the spherical harmonic and Fourier-plane formalisms. Estimators (\ref{eq:fullskyestimator}) separate the set of possible arguments $\bc{\ell_1,\ell_2,\ell_3}$ disjointly into parity even ($L$ even) and parity odd ($L$ odd), whereas $\hat{B}_+$ and $\hat{B}_-$ are defined on the same full set of angular frequency combinations\footnote{A similar behavior as for the spherical harmonic estimators would have been unexpected since the possible arguments of $\hat{B}_\pm$ form a non-countable set.}. In other words, when limiting $\hat{B}_+$ to integer angular frequencies only, the same information is contained in \lq half\rq\ the number of measures in the spherical harmonics case, namely those with $L$ even. The latter estimators have a covariance of half the size of the covariance of $\hat{B}_+$, so that the overall information content is the same for both approaches -- as required.

\subsection{Information content}
\label{sec:info}

We verify the findings of the foregoing section by comparing the information contained in both approaches in terms of the Fisher matrix \citep{tegmark97}. For a practical implementation we specialize to a non-tomographic weak lensing survey (see e.g. \citealp{bartelmann01} for an overview), assuming a cosmology-independent covariance that is well approximated by the Gaussian approximation, i.e. using (\ref{eq:bicovgauss}) and (\ref{eq:bicovtj}), respectively. To allow for direct comparison, we limit the Fourier-plane approach to integer $\ell$ with all bin sizes set to unity. Due to the symmetry under permutations of the arguments of the bispectra, one can impose the condition $\ell_1 \leq \ell_2 \leq \ell_3$ on both formalisms, rendering a block-wise diagonal covariance matrix. Inspecting (\ref{eq:bicovgauss}), the only dependence on the arguments of the second bispectrum, i.e. $\ell_4$ to $\ell_6$, is due to the Kronecker symbols (\ref{eq:kroneckers}), so that the summations over $\ell_4$ to $\ell_6$ become trivial. 

Hence, the Fisher matrix can be written as
\eq{
\label{eq:fisher}
F_{\mu\nu} = \sum_{l_{\rm min} \leq \ell_1 \leq \ell_2 \leq \ell_3 \leq l_{\rm max}} D_{\bar{\ell}_1,\bar{\ell}_2,\bar{\ell}_3,\bar{\ell}_1,\bar{\ell}_2,\bar{\ell}_3}\; \frac{\partial B(\ell_1,\ell_2,\ell_3)}{\partial p_\mu}\;  \frac{A\; \bar{\ell}_1 \bar{\ell}_2 \bar{\ell}_3\; \Delta \ell_1 \Delta \ell_2 \Delta \ell_3\; \Lambda \br{\bar{\ell}_1,\bar{\ell}_2,\bar{\ell}_3}}{(2\,\pi)^3\; P(\bar{\ell}_1) P(\bar{\ell}_2) P(\bar{\ell}_3)}\; \frac{\partial B(\ell_1,\ell_2,\ell_3)}{\partial p_\nu}\;,
}
where $D_{\bar{\ell}_1,\bar{\ell}_2,\bar{\ell}_3,\bar{\ell}_1,\bar{\ell}_2,\bar{\ell}_3}=6$ for equilateral triangles, $D_{\bar{\ell}_1,\bar{\ell}_2,\bar{\ell}_3,\bar{\ell}_1,\bar{\ell}_2,\bar{\ell}_3}=2$ for isosceles, and $D_{\bar{\ell}_1,\bar{\ell}_2,\bar{\ell}_3,\bar{\ell}_1,\bar{\ell}_2,\bar{\ell}_3}=1$ else. The derivatives are taken with respect to a set of cosmological parameters $\vek{p}$. In this toy example we use only the single parameter $\Omega_{\rm m}$, reducing the Fisher matrix to a scalar $F$. Besides, we restrict the angular frequency values to an unphysically small range between $l_{\rm min}=100$ and $l_{\rm max}=150$ for computational reasons.

Weak lensing power spectra are computed for a standard $\Lambda$CDM cosmology, including non-linear evolution via the fit formula of \citet{smith03}. The bispectra are obtained via perturbation theory \citep[e.g.][]{fry84}, using \citet{scoccimarro01} with the definition of the non-linear wave vector by \citet{takada04} to account for non-linear evolution. For the projections along the line of sight we assume a redshift probability distribution according to \citet{smail94} with $\beta=1.5$ and a deep survey of 0.9 median redshift. Shape noise is incorporated by replacing the power spectra in the covariances with
\eq{
\label{eq:psshapenoise}
\bar{P}(\ell) = P(\ell) + \frac{\sigma^2_\epsilon}{2 \bar{n}}\;,
}
where the ellipticity dispersion $\sigma_\epsilon=0.35$ and the galaxy number density $n=40\,\mbox{arcmin}^{-2}$ are set to typical values for planned space-based surveys. 

\begin{figure}[t]
\begin{minipage}[c]{.65\textwidth}
\centering
\includegraphics[scale=.33,angle=270]{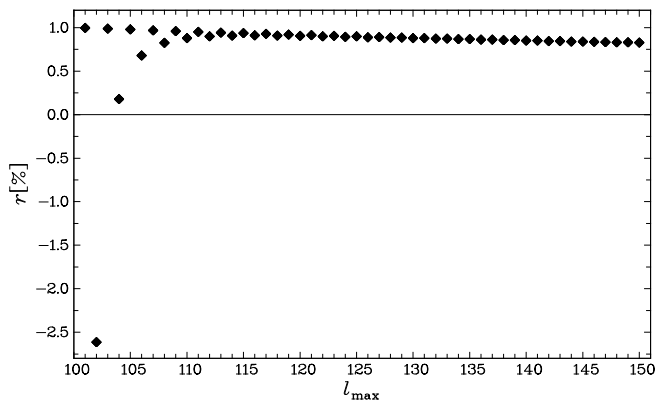}
\end{minipage}%
\begin{minipage}[c]{.35\textwidth}
\caption{Comparison of the Fisher information as obtained by spherical harmonics and Fourier-plane approach. Given is the relative deviation $r$ as a function of the maximum angular frequency $\ell_{\rm max}$ used in (\ref{eq:fisher}).}
\label{fig:compcov}
\end{minipage}
\end{figure}

We calculate the relative deviation of the Fisher information, $r \equiv F_{\rm Fourier}/F_{\rm sph.~harm.} -1$, as a function of $l_{\rm max}$. Note that, since we only consider ratios of $F$, the survey size $A$ drops out. Our results are shown in Fig.$\,$\ref{fig:compcov}. For $l_{\rm max}$ very close to $l_{\rm min}$ one sees alternating jumps in $r$ which can mostly be traced back to the fact that, due to the condition $\ell_1 \leq \ell_2 \leq \ell_3$, the terms entering (\ref{eq:fisher}) do not always split exactly half into $L$ even and odd. After this \lq burn in\rq\ for $l_{\rm max} \lesssim 120$, $r$ shows only little variation. The remaining offset from zero, which is slowly decreasing, can entirely be assigned to the different prefactors in the covariances, i.e. the terms related to the Wigner symbol and $\Lambda$, respectively. The range of angular frequencies plotted in Fig.$\,$\ref{fig:compcov} is still far from any physically relevant situation, but nonetheless the two approaches agree already better than $99\,\%$.

\section{Conclusions}
\label{sec:conclusions}

In this work we intended to give insight into the derivation and the form of the bispectrum covariance in the flat-sky approximation, based exclusively on the two-dimensional Fourier formalism. We defined an unbiased estimator that takes the average over the overlap of annuli in Fourier space, and computed its covariance. To obtain precise normalizations, a case distinction is necessary between degenerate and non-degenerate triangle configurations. However, given that both normalizations become very similar for $\ell \gg 1$, which is assumed in the flat-sky approach anyway, we suggest as a simple and fair approximation to use the expression derived for the degenerate case. Then our result for the Gaussian part of the bispectrum covariance reads
\eqa{
\label{eq:bicovgaussfinal}
{\rm Cov} \br{B(\bar{\ell}_1,\bar{\ell}_2,\bar{\ell}_3),\; B(\bar{\ell}_4,\bar{\ell}_5,\bar{\ell}_6)}_{\rm Gauss} = \frac{(2\,\pi)^3\; D_{\bar{\ell}_1,\bar{\ell}_2,\bar{\ell}_3,\bar{\ell}_4,\bar{\ell}_5,\bar{\ell}_6}}{A\; \bar{\ell}_1 \bar{\ell}_2 \bar{\ell}_3\; \Delta \ell_1 \Delta \ell_2 \Delta \ell_3 }\; \Lambda^{-1} \br{\bar{\ell}_1 + \frac{\Delta \ell_1}{2},\bar{\ell}_2 + \frac{\Delta \ell_2}{2},\bar{\ell}_3 + \frac{\Delta \ell_3}{2}}\; P(\bar{\ell}_1) P(\bar{\ell}_2) P(\bar{\ell}_3)\;.
}
This formula is readily generalized to the total covariance by modifying the arguments of $\Lambda$, appearing in the non-Gaussian terms of (\ref{eq:bicovtot}), accordingly. It is directly applicable to any real values of angular frequencies, to arbitrary binning, and to any compact, finite survey geometry. This formula can be modified to incorporate shot or shape noise, as well as to account for photometric redshift information or CMB polarization in a straightforward manner. 

While the general form of our result was in agreement with existing work, we found, contrary to \citet{hu00}, that the size of the covariance is a factor of 2 larger than the one obtained by the flat-sky spherical harmonic approach. By defining parity-sensitive bispectrum estimators, we discussed the behavior of both formalisms with respect to parity transformations, arguing that the difference in the covariances is indeed to be expected because in the spherical harmonic framework, parity-invariant measures are restricted to a subset of the angular frequency combinations at which the bispectra are evaluated. In a practical example we demonstrated that both approaches indeed contain the same information in terms of the Fisher matrix, with a high level of agreement. As a consequence, we can confirm that studies performed in the flat-sky spherical harmonic approach, such as \citet{takada04}, yield correct parameter constraints as long as the analysis is restricted to integer $\ell$ with the sum of the three angular frequencies being even.

We established a relation between the geometrical and intuitive process of averaging over the overlapping regions of annuli in the Fourier plane and the Wigner symbol of the spherical harmonic approach. Both quantities were demonstrated to be in turn connected to a simple measure that is proportional to the size of the area enclosed by the triangle configuration for which the bispectrum is calculated. This resulted in convenient, yet precise approximation formulae for the prefactors of the covariances of both approaches at consideration.

Under the assumption of a compact survey geometry and scales much smaller than the extent of the survey area, (\ref{eq:bicovtot}) provides a cleanly derived bispectrum covariance matrix that naturally incorporates the scaling with survey size, is not restricted to integer angular frequencies, and allows for any appropriate binning.

\begin{acknowledgements}
The authors would like to thank Bhuvnesh Jain and Masahiro Takada for helpful discussions and the referee for a helpful report. We thank Joel Berg{\'e} for comparison tests of our bispectrum codes. BJ is grateful to Sarah Bridle for kind hospitality at UCL. BJ acknowledges support by the Deutsche Telekom Stiftung and the Bonn-Cologne Graduate School of Physics and Astronomy. XS is supported by the International Max Planck Research School (IMPRS) for Astronomy and Astrophysics. This work was supported by the DFG under the Priority Programme 1177 `Galaxy Evolution' and within the Transregional Collaborative Research Centre TR33 `The Dark Universe'.
\end{acknowledgements}

\bibliographystyle{aa}

\end{document}